\documentclass[floats,aps,showpacs,twocolumn,longbibliography,superscriptaddress]{revtex4-1}

\usepackage[dvips]{graphicx}
\usepackage{amsfonts}
\usepackage{amsmath,amssymb}
\usepackage{dsfont}
\usepackage{placeins}
\usepackage{afterpage}
\usepackage{flafter}
\usepackage[usenames,dvipsnames]{color}

\newcommand{\map}{U}
\newcommand{\qmap}{\widehat{\map}}

\newcommand{\intindex}{\text{int}}
\newcommand{\mint}{m_{\text{\intindex}}}

\newcommand{\diag}{\text{diag}}
\newcommand{\inc}{\text{inc}}
\newcommand{\per}{\text{per}}

\newcommand{\act}{J}
\newcommand{\Arg}{\text{Arg}}

\newcommand{\K}{\kappa}                          % kicking strength of st. map
\newcommand{\Vrs}{V_{r:s}}
\newcommand{\Irs}{I_{r:s}}
\newcommand{\CanTrans}{\mathcal{T}}

\newcommand{\heff}{h}

\newcommand{\Uopen}{\widehat{\map}_{\text{o}}}
\newcommand{\UopenAlt}{\widehat{\map}'_{\text{o}}}
\newcommand{\pabs}{\widehat{P}_{\mathcal{L}}}

\newcommand{\oneOp}{\hat{\mathbf{1}}}
\newcommand{\qOp}{\hat{q}}
\newcommand{\pOp}{\hat{p}}
\newcommand{\myi}{i}

\newcommand{\ket}[1]{\left| #1 \right\rangle}

\newcommand{\braket}[2]{\left\langle #1| #2 \right\rangle}
\newcommand{\braOpket}[3]{\langle #1 | #2 | #3 \rangle}

\newcommand{\Ket}[1]{| #1 \rangle}

\newcommand{\BraKet}[2]{\langle #1| #2 \rangle}
\newcommand{\BraOpKet}[3]{\langle #1 | #2 | #3 \rangle}

\makeatletter
\let\Hy@backout\@gobble
\makeatother

\usepackage{color}

\begin{document}

\title{Perturbation-Free Prediction of Resonance-Assisted Tunneling \\ in Mixed
Regular--Chaotic Systems}

\author{Normann Mertig}
\affiliation{Technische Universit\"at Dresden, Institut f\"ur Theoretische
             Physik and Center for Dynamics, 01062 Dresden, Germany}
\affiliation{Max-Planck-Institut f\"ur Physik komplexer Systeme, N\"othnitzer
Stra\ss{}e 38, 01187 Dresden, Germany}
\affiliation{Department of Physics, Tokyo Metropolitan University,
Minami-Osawa, Hachioji 192-0397, Japan}

\author{Julius Kullig}
\affiliation{Technische Universit\"at Dresden, Institut f\"ur Theoretische
             Physik and Center for Dynamics, 01062 Dresden, Germany}
\affiliation{Max-Planck-Institut f\"ur Physik komplexer Systeme, N\"othnitzer
Stra\ss{}e 38, 01187 Dresden, Germany}
\affiliation{Institut f\"ur Theoretische Physik, 
Otto-von-Guericke-Universit\"at Magdeburg,
Postfach 4120, 39016 Magdeburg, Germany}

\author{Clemens L\"obner}
\affiliation{Technische Universit\"at Dresden, Institut f\"ur Theoretische
             Physik and Center for Dynamics, 01062 Dresden, Germany}
\affiliation{Max-Planck-Institut f\"ur Physik komplexer Systeme,
N\"othnitzer Stra\ss{}e 38, 01187 Dresden, Germany}

\author{Arnd B\"acker}\
\affiliation{Technische Universit\"at Dresden, Institut f\"ur Theoretische
             Physik and Center for Dynamics, 01062 Dresden, Germany}
\affiliation{Max-Planck-Institut f\"ur Physik komplexer Systeme, N\"othnitzer
Stra\ss{}e 38, 01187 Dresden, Germany}

\author{Roland Ketzmerick}
\affiliation{Technische Universit\"at Dresden, Institut f\"ur Theoretische
             Physik and Center for Dynamics, 01062 Dresden, Germany}
\affiliation{Max-Planck-Institut f\"ur Physik komplexer Systeme, N\"othnitzer
Stra\ss{}e 38, 01187 Dresden, Germany}

\date{\today}

\begin{abstract}

For generic Hamiltonian systems we derive predictions for dynamical tunneling
from regular to chaotic phase-space regions.
In contrast to previous approaches, we account for the resonance-assisted
enhancement of regular-to-chaotic tunneling in a non-perturbative way.
This provides the foundation for future semiclassical complex-path evaluations
of resonance-assisted regular-to-chaotic tunneling.
Our approach is based on a new class of integrable approximations which mimic
the regular phase-space region and its dominant nonlinear resonance chain in a
mixed regular--chaotic system.
We illustrate the method for the standard map.

\end{abstract}

\maketitle
\noindent

\section{Introduction}

Tunneling is a fundamental effect in wave mechanics, which allows for entering
classically inaccessible regions.
While textbooks focus on tunneling through potential barriers, tunneling
processes in nature often take place in the absence of such energetic barriers.
Instead one observes dynamical tunneling \cite{DavHel1981, KesSch2011}
between classically disjoint regions in phase space.

\begin{figure}[b!]
  \begin{center}
    \includegraphics[]{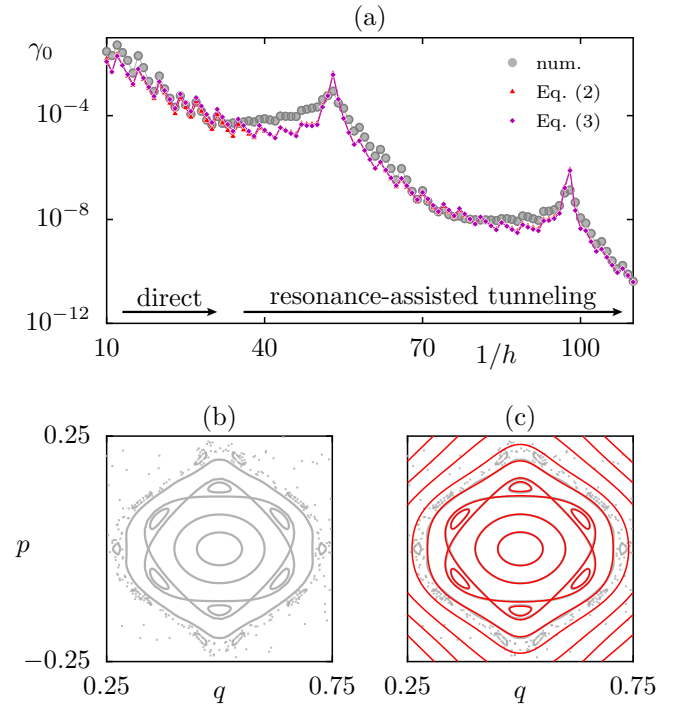}
     \caption{(color online) (a) Regular-to-chaotic decay rate $\gamma_{0}$
 versus $1/\heff$ for the standard map at $\K=3.4$. The numerically determined
rates (gray dots) are compared to (the sum of incoherent terms of) the
predictions of Eq.~\eqref{eq:GammaPrediction} ([red] triangles) and
Eq.~\eqref{eq:GammaPredictionAlt} ([magenta] squares). (b) Phase space with
regular orbits (lines) and a chaotic orbit (dots) including a 6:2 nonlinear
resonance chain. (c) Like (b) with an integrable approximation ([red] lines) on
top.}
     \label{fig:RAT_Intro}
  \end{center}
\end{figure}

In generic Hamiltonian systems dynamical tunneling usually occurs between
regions of regular and chaotic motion.
For a typical phase space of a mixed regular--chaotic system see
Fig.~\ref{fig:RAT_Intro}(b).
In particular, while a classical particle cannot traverse from the regular
to the chaotic region, a wave can tunnel from the regular to the chaotic region.
This regular-to-chaotic tunneling process manifests itself impressively in
chaos-assisted tunneling \cite{LinBal1990, BohTomUll1993}.

Until today the importance of regular-to-chaotic tunneling has been
demonstrated in numerous experiments, including optical microcavities
\cite{ShiHarFukHenSasNar2010, ShiHarFukHenSunNar2011,
YanLeeMooLeeKimDaoLeeAn2010, KwaShiMooLeeYanAn2015, YiYuLeeKim2015, 
YiYuKim2016}, microwave billiards
\cite{DemGraHeiHofRehRic2000, BaeKetLoeRobVidHoeKuhSto2008, DieGuhGutMisRic2014,
GehLoeShiBaeKetKuhSto2015}, and cold atom systems \cite{Hen2001,
SteOskRai2001}.
A recent success being the experimental verification
\cite{KwaShiMooLeeYanAn2015, GehLoeShiBaeKetKuhSto2015} that tiny nonlinear
resonance chains within the regular region, as shown in
Fig.~\ref{fig:RAT_Intro}(b), indeed drastically enhance tunneling as predicted
in Refs.~\cite{UzeNoiMar1983, Ozo1984, BroSchUll2001, BroSchUll2002}.
Furthermore, regular-to-chaotic tunneling is expected to play an important role
for atoms and molecules in strong fields, as discussed in
Refs.~\cite{ZakDelBuc1998, BucDelZak2002, WimSchEltBuc2006}.

Motivated by these applications regular-to-chaotic tunneling is also a field of
intense theoretical research \cite{ShuIke1995, ShuIke1998, PodNar2003,
PodNar2005, EltSch2005, WimSchEltBuc2006, SheFisGuaReb2006, BaeKetLoeSch2008,
BaeKetLoeRobVidHoeKuhSto2008, ShuIshIke2008, ShuIshIke2009a, ShuIshIke2009b,
BaeKetLoeWieHen2009, BaeKetLoe2010, LoeBaeKetSch2010, MerLoeBaeKetShu2013,
HanShuIke2015, KulWie2016}, which is mainly focused on periodically driven 
model systems with one degree of freedom.
Here, a major achievement is the combination of (i) direct
\cite{BaeKetLoeSch2008, BaeKetLoe2010} and (ii) resonance-assisted
\cite{UzeNoiMar1983, Ozo1984, BroSchUll2001, BroSchUll2002} regular-to-chaotic
tunneling in a single prediction \cite{LoeBaeKetSch2010, SchMouUll2011}.
This prediction shows that as function of decreasing
effective Planck's constant $\heff$ one has two corresponding regimes:
(i) Regular states localize on a single quantizing torus.
In this regime, tunneling is determined by direct transitions from this regular
torus into the chaotic region \cite{BaeKetLoeSch2008, BaeKetLoe2010}
which can be evaluated semiclassically using complex paths
\cite{MerLoeBaeKetShu2013}.
(ii) For even smaller $\heff$ a regular state, while still mostly concentrated
on the main quantizing torus, acquires resonance-assisted contributions
on further quantizing tori \cite{UzeNoiMar1983, Ozo1984, BroSchUll2001,
BroSchUll2002} located more closely to the border of the regular region,
see Fig.~\ref{fig:Istates}(c) for an illustration.
This resonance-assisted contribution dominates
tunneling to the chaotic region \cite{EltSch2005, LoeBaeKetSch2010,
SchMouUll2011}.
Thus, one observes a resonance-assisted enhancement of regular-to-chaotic
tunneling.
For an example of this enhancement see Fig.~\ref{fig:RAT_Intro}(a).
Note, that for much smaller $\heff$ there is even a third regime for which
regular states may localize within the resonance chain. This regime is not
considered here.

Despite the above achievements, a semiclassical evaluation of resonance-assisted
tunneling in mixed regular--chaotic systems remains an open problem.
In particular, the state of the art predictions \cite{LoeBaeKetSch2010,
SchMouUll2011} defy a semiclassical evaluation using the techniques developed
for integrable systems~\cite{DeuMou2010, DeuMouSch2013}.
More specifically, so far (i) an integrable approximation of the regular region
which ignores resonance chains is used to predict the magnitude of direct
tunneling transitions from quantizing tori towards the chaotic region
\cite{BaeKetLoeSch2008, BaeKetLoe2010}.
Subsequently, (ii) resonance-assisted contributions are taken into account
by perturbatively solving \cite{BroSchUll2002} an additional pendulum
Hamiltonian which models the relevant resonance chain \cite{LoeBaeKetSch2010,
SchMouUll2011}.
However, only a perturbation-free prediction, based on a single integrable
approximation which includes the relevant resonance chain will allow for a
semiclassical evaluation of resonance-assisted regular-to-chaotic tunneling in
the spirit of Refs.~\cite{DeuMou2010, DeuMouSch2013}.

In this paper we derive such perturbation-free predictions of resonance-assisted
regular-to-chaotic tunneling.
They are based on a new class of integrable approximations $H_{r:s}$
\cite{KulLoeMerBaeKet2014} which include the dominant $r$:$s$ resonance, see
Fig.~\ref{fig:RAT_Intro}(c).
In particular, the eigenvalue equation
\begin{align}
   \label{eq:HrsEigenvalue}
  \widehat{H}_{r:s} \ket{\mint} = E_m \ket{\mint},
\end{align}
of such integrable approximations $H_{r:s}$ provides eigenstates $\ket{\mint}$
which model the localization of regular states on the regular phase-space
region, explicitly including the resonance-assisted contributions on multiple
quantizing tori, in a non-perturbative way.
Using such states allows for extending the results of
Refs.~\cite{BaeKetLoeSch2008, BaeKetLoe2010} to the case of resonance-assisted
tunneling.

In particular, the decay rates $\gamma_m$ of metastable states which localize on
the regular phase-space region and decay via regular-to-chaotic tunneling can be
predicted according to
\begin{align}
   \label{eq:GammaPrediction}
  \gamma_{m} \approx \Gamma_{m}(t=1) := \big\|\pabs \qmap \ket{\mint}
\big\|^2.
\end{align}
Here $\qmap$ is the time evolution operator and $\pabs$ is a projector onto a
leaky region $\mathcal{L}$ located in the chaotic part of phase space.

We further show that regular-to-chaotic decay rates can be predicted with
similar accuracy, when using a simplified formula which no longer contains the
time-evolution operator.
Instead it evaluates only the probability of the state $\ket{\mint}$ on the
leaky region $\mathcal{L}$
\begin{align}
   \label{eq:GammaPredictionAlt}
  \gamma_{m} \approx  \Gamma_{m}(t=0) := \big\|\pabs \ket{\mint} \big\|^2.
\end{align}
Both perturbation-free predictions, Eqs.~\eqref{eq:GammaPrediction} and
\eqref{eq:GammaPredictionAlt}, give good results for the standard map,
see Fig.~\ref{fig:RAT_Intro}.
In that both predictions provide the foundation for future semiclassical
predictions of resonance-assisted regular-to-chaotic tunneling
\cite{FriMerLoeBaeKet}.
We remark that the prediction of Eq.~\eqref{eq:GammaPrediction} has previously
been evaluated semiclassically for integrable approximations without resonances
\cite{MerLoeBaeKetShu2013}, using the time-domain techniques of
Refs.~\cite{ShuIke1995, ShuIke1998, ShuIshIke2008, ShuIshIke2009a,
ShuIshIke2009b} giving predictions for direct regular-to-chaotic tunneling.
However, we believe that a future semiclassical prediction of resonance-assisted
regular-to-chaotic tunneling would be easier obtained from
Eq.~\eqref{eq:GammaPredictionAlt}, since it does not involve any time
evolution and thus allows for a semiclassical evaluation using the simpler
WKB-like techniques of Refs.~\cite{DeuMou2010, DeuMouSch2013}.

The paper is organized as follows:\
In Sec.~\ref{Sec:ExampleSystem} we introduce the standard map as a paradigmatic
Hamiltonian example system with a mixed phase space.
We further present regular-to-chaotic decay rates as a measure of
regular-to-chaotic tunneling and discuss their numerical evaluation.
In Sec.~\ref{Sec:RAT} we derive the predictions,
Eqs.~\eqref{eq:GammaPrediction} and \eqref{eq:GammaPredictionAlt}.
In Sec.~\ref{Sec:ResultsStandardMap} we illustrate how these predictions are
evaluated using the example of the standard map.
In Sec.~\ref{Sec:Results} we present our results and compare them to the
perturbative predictions of Refs.~\cite{LoeBaeKetSch2010, SchMouUll2011}.
In Sec.~\ref{Sec:Discussion} we discuss the main approximations and
limitations of our approach.
A summary and outlook is given in Sec.~\ref{Sec:Summary}.

\section{Example System}
  \label{Sec:ExampleSystem}

In this paper we focus on periodically driven Hamiltonian systems with one
degree of freedom, which exhibit all generic features of a mixed phase space.
Classically, the stroboscopic map
\begin{align}
  \label{eq:StroboscopicMap}
  \map: (q_n, p_n) \mapsto (q_{n+1}, p_{n+1}),
\end{align}
describes the evolution of positions and momenta, $(q,p)$, in phase space
from time $t=n$ to $t=n+1$ over one period of the external driving.
Quantum-mechanically, the time-evolution is given by the corresponding
unitary time-evolution operator $\qmap$.

In Sec.~\ref{Sec:StandardMap} we introduce the standard map as a paradigmatic
example of a periodically driven one-degree-of-freedom system with a mixed phase
space.
In Sec.~\ref{Sec:TunnelingRatesInTheStandardMap} we introduce regular-to-chaotic
decay rates $\gamma$, as the central object of our investigation.
Furthermore, we discuss their numerical computation.
Particular attention is paid to nonlinear resonance chains and their quantum
manifestations.

\subsection{Standard Map}
  \label{Sec:StandardMap}

Classically, the standard map originates from a periodically kicked Hamiltonian
with one degree of freedom $H(q,p,t) = T(p) + V(q) \sum_{n\in\mathbb{N}}
\delta(n-t)$.
Here, $\delta(\cdot)$ is the Dirac delta function.
For the standard map $T(p)=p^2/2$ and $V(q) = \K/(2\pi)^2\cos{(2\pi q)}$, where
$\K$ is the kicking strength.
Its stroboscopic map $\map$ \cite{Chi1979}, Eq.~\eqref{eq:StroboscopicMap}, in
its symmetrized version is given by
\begin{subequations}
  \label{eq:SMap}
  \begin{align}
	q_{n+1} =&\, q_n + p_n + \frac{\K}{4\pi} \sin(2\pi q_n), \\
	p_{n+1} =&\, p_n  + \frac{\K}{4\pi} \sin(2\pi q_n) + \frac{\K}{4\pi}
\sin(2\pi q_{n+1}),
  \end{align}
\end{subequations}
where $(q_n, p_n)$ represents a phase-space point in the middle of the $n$th
kick.
For convenience the standard map is considered on a torus $(q, p) \in [0,1[
\times [-0.5, 0.5[$ with periodic boundary conditions.

In this paper we mainly focus on kicking strength $\K=3.4$.
Here, the phase space exhibits a large regular region which is centered around 
an elliptic fixed point, see Fig.~\ref{fig:RAT_Intro}(b).
As expected from the KAM theorem \cite{Kol1954, Arn1963, Arn1963b, Mos1962} the
regular region consists of one-dimensional invariant tori.
Along these tori orbits of regular motion rotate around the fixed point.
These tori are interspersed by nonlinear resonance chains, wherever $s$
rotations of a regular orbit match $r$ periods of the external driving
\cite{Bir1913, LicLie1983a, Chi1979}.
For example, the standard map at $\kappa=3.4$ has a dominant $r$:$s=6$:$2$
resonance, leading to the six regular sub-regions in
Fig.~\ref{fig:RAT_Intro}(b).
Note that we choose the numbers $r$ and $s$ in the ratio $r$:$s$ such that
$r$ is the number of sub-regions of the resonance.
The region of regular motion is embedded in a region of chaotic motion.

The quantum-mechanical analogue of the stroboscopic map $\map$ is the unitary
time-evolution operator \cite{BerBalTabVor1979, HanBer1980, ChaShi1986,
KeaMezRob1999, DegGra2003b, Bae2003}
\begin{align}
  \label{eq:Qmap}
 \hspace*{-0.2cm} \qmap = \exp\left(-\myi\frac{V(\qOp)}{2\hbar}\right)
\exp\left(-\myi\frac{T(\pOp)}{\hbar} \right)
\exp\left(-\myi\frac{V(\qOp)}{2\hbar}\right).
\end{align}
Here, $\heff = 2\pi\hbar$ is the effective Planck constant and $\qOp$ and $\pOp$
are the operators of position and momentum, respectively.
Similar to the classical case we consider $\qmap$ on a toric phase space, which
leads to grids in position and momentum space \cite{BerBalTabVor1979,
HanBer1980, ChaShi1986, KeaMezRob1999, DegGra2003b, Bae2003}
\begin{subequations}
\begin{align}
\label{eq:qnDef}
\overline{q}_n &= \heff (n+\theta_p),\quad \text{with}\quad
\overline{q}_{n}\in[0,1[\\
\label{eq:pnDef}
\overline{p}_n &= \heff (n+\theta_q),\quad \,\text{with}\quad
\overline{p}_{n}\in[-0.5,0.5[,
\end{align}
\end{subequations}
with $n\in \mathbb{N}$.
This implies that the inverse of the effective Planck constant is a natural
number $1/\heff=N \in \mathbb{N}$, giving the dimension of the Hilbert space.
For the standard map, we choose the Bloch phase $\theta_p=0$, while $\theta_q=0$
if $N$ is even and $\theta_q=0.5$ if $N$ is odd.
This gives the finite-dimensional time-evolution operator in position
representation
\begin{align}
  \label{eq:StandardQmapOnTorus}
  \braOpket{\overline{q}_n}{\qmap}{\overline{q}_k} \! &= \! \\ \nonumber
  & \hspace*{-1.0cm}\frac{e^{-\myi\pi/4}}{\sqrt{N}}
\exp\!\left(\myi \,2\pi N \!
\left[- \frac{V(\overline{q}_n)}{2} +
\frac{(\overline{q}_n-\overline{q}_k)^2}{2}
- \frac{V(\overline{q}_k)}{2}\right]\!\right),
\end{align}
with $n,k = 0, \dots ,N-1$.

In the following it is fundamental that eigenstates of a
mixed regular--chaotic system can be
classified according to their semiclassical localization on the regular or
chaotic region, respectively.
More specifically, chaotic states spread across the chaotic region
\cite{Per1973, Ber1977b, Vor1979}, while regular states localize on a torus
$\tau_{m}$ of the regular region which has quantizing action \cite{Boh1913,
Boh1913b, Som1916}
\begin{align}
  \label{eq:BohrSommerfeld}
  \act_{m} := \frac{1}{2\pi}\oint_{\tau_{m}} p(q)\: \text{d}q
        = (m + 1/2) \hbar,
\end{align}
labeled by an index $m\in\mathbb{N}$.
In order to account for resonance-assisted tunneling it is further indispensable
to consider the finer structure of regular states.
In particular, it will be crucial that a regular state $m$ localizes not only on a
dominant quantizing torus $\act_m$.
Instead, an $r$:$s$ resonance induces additional contributions on the tori
$\act_{m+kr}$ with $k\in\mathbb{Z}$, see Refs.~\cite{UzeNoiMar1983, Ozo1984,
BroSchUll2001, BroSchUll2002, WisSarArrBenBor2011, Wis2014, WisSch2015} and 
references therein.

\subsection{Regular-to-Chaotic Decay Rates in the Standard Map}
  \label{Sec:TunnelingRatesInTheStandardMap}

In this section we introduce regular-to-chaotic decay rates $\gamma$ of an open
system for quantifying regular-to-chaotic tunneling.
Note that in closed systems chaos-assisted tunnel splittings
\cite{BohTomUll1993} are an often-used alternative \cite{BaeKetLoe2010}.

Our general approach for defining regular-to-chaotic decay rates proceeds in
three steps:\
(a) We introduce a leaky region $\mathcal{L}$ within the chaotic part of phase
space,
(b) we determine the \emph{decay rates} of its regular states, and
(c) we classify the corresponding decay rates as \emph{regular-to-chaotic
decay rates}.
Step (c) is justified because each regular state of the open system decays by
regular-to-chaotic tunneling towards the chaotic region and subsequently
entering the leaky region within the chaotic part of phase space.

More specifically, we proceed by (a) introducing a projector $\pabs$ which
absorbs probability on a phase-space region $\mathcal{L}$ within the chaotic
part of phase space.
Based on this projector and the unitary time-evolution operator $\qmap$ of the
closed system we define the time-evolution operator of the open system as
\begin{align}
  \label{eq:UopenNum}
  \Uopen = (\oneOp - \pabs)\hat{U}(\oneOp - \pabs).
\end{align}
(b) We solve its eigenvalue equation
\begin{equation}
  \label{eq:UopenEigenvalueNum}
  \Uopen\ket{m} = \exp\left(i\phi_m-\frac{\gamma_m}{2}\right)\ket{m}.
\end{equation}
Here, $\ket{m}$ represents a metastable, right eigenvector of the sub-unitary
operator $\Uopen$.
The corresponding eigenvalue is determined by an eigenphase $\phi_m$ and a
decay rate $\gamma_{m}$.
The latter describes the exponential decay of $\ket{m}$ in time.
(c) We assign to each regular state $\ket{m}$ a label $m$ according to its
dominant localization on the quantizing torus $\act_m$ and refer to its decay
rate $\gamma_m$ as the regular-to-chaotic decay rate.

Specifically, for the standard map (a) we use
\begin{align}
 \label{eq:LeakyRegion}
  \!\!\mathcal{L} := \left\{(q,p) \; \left|\right. \;\; q <
q_l\;\;\;\text{or}\;\;\;q > q_r := 1- q_l \right\},
\end{align}
and define the projector
\begin{align}
  \label{eq:Pabs}
  \pabs \ket{q} = \chi(q)\ket{q} \quad \text{with }
   \chi(q) = \left\{ \begin{array}{l l}
                   1 & \text{for } (q, \cdot) \in \mathcal{L}\\
                   0 & \text{for } (q, \cdot) \notin \mathcal{L}
                   \end{array}
           \right. .
\end{align}
Here, we choose $q_l$ close to the regular--chaotic border.
This ensures that $\gamma_m$, which depends on the choice of the leaky region
$\mathcal{L}$, is dominated by tunneling from the regular towards the chaotic
region.
For a more detailed discussion see
Sec.~\ref{Sec:BeyondRTCTunneling}.
(b) We compute the finite-dimensional matrix representation of $\Uopen$ for each
value of $1/\heff\in\mathbb{N}$.
To this end we set all those entries in Eq.~\eqref{eq:StandardQmapOnTorus}
equal to zero, for which either $\overline{q}_{n}$ or $\overline{q}_{k}$ are in
the leaky region $\mathcal{L}$.
We diagonalize the resulting $\Uopen$ numerically.
(c) The regular-to-chaotic decay rates $\gamma_{m}$ are labeled according
to the dominant localization of $\ket{m}$ on the quantizing tori $\act_m = \hbar
(m + 1/2)$.

We present the numerically obtained regular-to-chaotic decay rates
$\gamma_{0}$ of the standard map at $\kappa=3.4$ as a function of the inverse
effective Planck constant ([gray] dots) in Fig.~\ref{fig:RAT_Intro}(a).
The numerical results are consistent with the expectations due to
Refs.~\cite{LoeBaeKetSch2010, SchMouUll2011}:\
(i) For $1/ \heff \lesssim 35$ the state $\ket{0}$ localizes on the torus
$\act_0$ such that the direct tunneling from $\act_0$ to $\mathcal{L}$
dominates.
In this regime, $\gamma_0$ decreases exponentially for decreasing $\heff$ which
is a characteristic feature of direct transitions, see Ref.~\cite{HanOttAnt1984,
BaeKetLoeSch2008, BaeKetLoe2010, MerLoeBaeKetShu2013}.
(ii) In the regime $1/ \heff \gtrsim 35$ tunneling is enhanced by the $6$:$2$
resonance.
For $35 \lesssim 1/\heff \lesssim 80$ the resonance contribution of the state
$\ket{0}$ on $\act_{6}$ is significant such that direct tunneling transition
from $\act_{6}$ to $\mathcal{L}$ dominates $\gamma_m$.
This leads to a peak at $1/\heff=53$, where the state $\ket{0}$ has half its
weight on $\act_{6}$.
Finally, for $1/\heff\gtrsim 80$ the resonance contribution of $\ket{0}$ on
$\act_{12}$ is significant such that direct tunneling from $\act_{12}$ to
$\mathcal{L}$ dominates the decay rate $\gamma_m$, with a peak at
$1/\heff=98$.
In Fig.~\ref{fig:Results}(a,c) we show similar numerical rates ([gray] dots)
for the standard map at $\K=2.9$ and $\K=3.5$ with a dominating $10$:$3$ and
$6$:$2$ resonance, respectively.

\section{Perturbation-Free Predictions of Resonance-Assisted Regular-to-chaotic
Tunneling}
  \label{Sec:RAT}

In this section we derive the perturbation-free predictions for
resonance-assisted regular-to-chaotic decay rates.
In Sec.~\ref{Sec:DerivationWithTimeEvolution} we derive
Eq.~\eqref{eq:GammaPrediction} which uses the time-evolution operator.
In Sec.~\ref{Sec:DerivationWithoutTimeEvolution} we derive
Eq.~\eqref{eq:GammaPredictionAlt} which does not use the time-evolution
operator.

\subsection{Derivation of  Eq.~\eqref{eq:GammaPrediction} with Time Evolution}
  \label{Sec:DerivationWithTimeEvolution}

The starting point for deriving Eq.~\eqref{eq:GammaPrediction} is the definition
of the regular-to-chaotic decay rate $\gamma_m$ from the appropriate eigenvalue
problem.
We use the same definitions as for the numerical determination of
regular-to-chaotic decay rates, see Eqs.~\eqref{eq:UopenNum} and
\eqref{eq:UopenEigenvalueNum} of Sec.~\ref{Sec:TunnelingRatesInTheStandardMap}.
They are repeated for convenience, namely a general sub-unitary operator
\begin{align}
  \label{eq:Uopen}
  \Uopen &:= (\oneOp - \pabs)\hat{U}(\oneOp - \pabs),
\end{align}
and its eigenvalue equation
\begin{align}
  \label{eq:UopenEigenvalue}
  \Uopen\ket{m} &= \exp\left(i\phi_m-\frac{\gamma_m}{2}\right)\ket{m}.
\end{align}
Here, the unitary operator $\qmap$ describes the time evolution of a mixed
regular--chaotic system over one unit of time.
Furthermore, $\pabs$ is a projection operator which absorbs probability on the
leaky region $\mathcal{L}$ within the chaotic part of phase space.

For decay rates of such systems, it can be shown, that the following formula
applies, see App.~\ref{App:Derivation} for details,
\begin{equation}
  \label{eq:GammaPredOpen}
  \gamma_{m} = - \log\left(1 - \big\|\pabs \qmap \!\ket{m}
\big\|^2\right)\!
\stackrel{\gamma_m \ll 1}{\approx} \!\big\|\pabs \qmap \!\ket{m} \big\|^2,
\end{equation}
i.\,e., a regular-to-chaotic decay rate $\gamma_m$ (for which $\gamma_m\ll1$) is
given by the probability transfer from the regular state $\ket{m}$ into the
leaky region $\mathcal{L}$ via the unitary time-evolution operator $\qmap$.
Equation~\eqref{eq:GammaPredOpen} is as such not useful, since it still contains
the unknown eigenvector $\ket{m}$.
In particular, it would require to solve Eq.~\eqref{eq:UopenEigenvalue} which defines
$\gamma_m$ in the first place.
Hence, we proceed in the spirit of Refs.~\cite{BaeKetLoeSch2008, BaeKetLoe2010},
i.\,e., we approximate $\ket{m}$ using the eigenstates $\ket{\mint}$ of an
integrable approximation $H_{r:s}$, leading to our prediction
Eq.~\eqref{eq:GammaPrediction}.

The novel point of this paper is the use of an integrable approximation
$H_{r:s}$, which includes the dominant $r$:$s$ resonance.
This ensures that $\ket{\mint}$ models not only the localization of $\ket{m}$
on the main quantizing torus $\act_m$ but also accounts for the
resonance-assisted contributions on the tori $\act_{m+kr}$.
Precisely this extends Eq.~\eqref{eq:GammaPrediction}, as previously used in
\cite{BaeKetLoeSch2008, BaeKetLoe2010} for direct tunneling, to the regime
of resonance-assisted regular-to-chaotic tunneling in a non-perturbative way.

\subsection{Derivation of Eq.~\eqref{eq:GammaPredictionAlt} without Time Evolution}
  \label{Sec:DerivationWithoutTimeEvolution}

In this section we derive Eq.~\eqref{eq:GammaPredictionAlt}.
It predicts regular-to-chaotic decay rates from the localization of the mode
$\ket{\mint}$ on the leaky region $\mathcal{L}$.
In contrast to Eq.~\eqref{eq:GammaPrediction} it does not use the
time-evolution operator.
In that, Eq.~\eqref{eq:GammaPredictionAlt} is an ideal starting point for future
semiclassical predictions of regular-to-chaotic decay rates
\cite{FriMerLoeBaeKet} in the spirit of Refs.~\cite{DeuMou2010, DeuMouSch2013}.
In particular, it avoids the complications which arise in a semiclassical
evaluation of Eq.~\eqref{eq:GammaPrediction} due to the time-evolution operator.
We further remark that predictions like Eq.~\eqref{eq:GammaPredictionAlt} are
common for open systems.
For regular-to-chaotic decay rates they have heuristically been used,
e.\,g. in Refs.~\cite{BroSchUll2002, PodNar2003, PodNar2005, SchMouUll2011}.
Here, the main purpose of deriving Eq.~\eqref{eq:GammaPredictionAlt} is to
explicitly point out the involved approximations.

The derivation starts from an alternative definition of the sub-unitary
time-evolution operator
\begin{align}
 \label{eq:UopenAlt}
 \UopenAlt := \qmap (\oneOp - \pabs),
\end{align}
which satisfies the eigenvalue equation
\begin{align}
 \label{eq:UopenAltEigenvalue}
 \UopenAlt\ket{m'} = \exp\left(i\phi_{m}-\frac{\gamma_{m}}{2}\right)\ket{m'}.
\end{align}
Compare with Eqs.~\eqref{eq:Uopen} and \eqref{eq:UopenEigenvalue}.
As shown in App.~\ref{App:Isospectrality} the operators $\Uopen$ and
$\UopenAlt$ are isospectral.
Therefore, they exhibit the same eigenvalues, which give rise to the same
regular-to-chaotic decay rates $\gamma_m$.
Furthermore, the corresponding normalized right eigenvectors can be transformed
into each other, see App.~\ref{App:Isospectrality}.
We find,
\begin{align}
  \label{eq:SameLocalization}
 \ket{m} =
\frac{1}{\exp\left(i\phi_{m}-\frac{\gamma_{m}}{2}\right)}(\oneOp-\pabs)\ket{m'},
\end{align}
which implies that $\ket{m}$ and $\ket{m'}$ localize on the quantizing tori
$\act_{m+kr}$ of the regular region with equal probability (for $\gamma_m\ll1$).
On the other hand, $\ket{m'}$ is the time-evolved mode $\ket{m}$ according to
\begin{align}
  \label{eq:TimeEvolvedMode}
 \ket{m'} = \qmap \ket{m}.
\end{align}
Inserting Eq.~\eqref{eq:TimeEvolvedMode} into Eq.~\eqref{eq:GammaPredOpen} gives
\begin{align}
  \label{eq:GammaPredOpenAlt}
  \gamma_{m} = - \log\left(1 - \big\|\pabs \ket{m'} \big\|^2\right)
\stackrel{\gamma_{m}\ll1}{\approx} \big\|\pabs \ket{m'} \big\|^2,
\end{align}
which shows that a regular-to-chaotic decay rate $\gamma_m$ (for which
$\gamma_m\ll1$) is equivalent to the probability to find $\ket{m'}$ on the leaky
region $\mathcal{L}$.
Similar to Eq.~\eqref{eq:GammaPredOpen}, Eq.~\eqref{eq:GammaPredOpenAlt} is as such not
helpful, because it still contains the eigenvector $\ket{m'}$.
In particular, it would require to solve Eq.~\eqref{eq:UopenAltEigenvalue} which
defines $\gamma_m$ in the first place.
Hence, we approximate the mode $\ket{m'}$ using the more accessible eigenstates
$\ket{\mint}$ of an integrable approximation $H_{r:s}$, leading to our
prediction Eq.~\eqref{eq:GammaPredictionAlt}.

Here, the key point is again the use of integrable approximations $H_{r:s}$
which includes the relevant $r$:$s$ resonance.
Therefore, $\ket{\mint}$ models not only the localization of $\ket{m'}$ on
the main quantizing torus $\act_m$ but also its resonance-assisted contributions
on the tori $\act_{m+kr}$.
Precisely this allows for predicting resonance enhanced regular-to-chaotic
decay rates from Eq.~\eqref{eq:GammaPredictionAlt} in a non-perturbative
way.

An application of the predictions, Eqs.~\eqref{eq:GammaPrediction} and
\eqref{eq:GammaPredictionAlt}, for the standard map is demonstrated in
Sec.~\ref{Sec:ResultsStandardMap}.
The key approximation, i.\,e., modeling metastable regular states $\ket{m}$ (or
$\ket{m'}$) in terms of eigenstates $\ket{\mint}$ of an integrable approximation
$H_{r:s}$, is discussed in Sec.~\ref{Sec:DiscussionStates}.
Moreover, a comparison of the non-perturbative predictions,
Eqs.~\eqref{eq:GammaPrediction} and \eqref{eq:GammaPredictionAlt}, to the
perturbative predictions of Refs.~\cite{LoeBaeKetSch2010,
SchMouUll2011} is given in Sec.~\ref{Sec:ResultsPerturbation}.

\section{Perturbation-free Prediction of Tunneling in the Standard Map}
  \label{Sec:ResultsStandardMap}

In this section we illustrate our approach by applying it to the standard map.
In Sec.~\ref{Sec:DominantResonance}, we determine the $r$:$s$ resonance which
dominates tunneling.
In Sec.~\ref{Sec:IntegrableApproximation}, we set up an integrable approximation
including the nonlinear resonance chain using the iterative canonical
transformation method \cite{LoeLoeBaeKet2013, KulLoeMerBaeKet2014} as presented
in Ref.~\cite{KulLoeMerBaeKet2014}.
In Sec.~\ref{Sec:Quantization}, we quantize the integrable approximation and
determine its eigenstates $\ket{\mint}$ from Eq.~\eqref{eq:HrsEigenvalue}.
Finally, the results will be discussed in the next section,
Sec.~\ref{Sec:Results}.

\subsection{Choosing the Relevant Resonance}
  \label{Sec:DominantResonance}

In order to apply our prediction it is crucial to first identify the $r$:$s$
resonance which dominates the tunneling process.
A detailed discussion as to which resonance dominates tunneling in which regime,
can be found in Ref.~\cite{LoeBaeKetSch2010}.
Here, we focus on the $r$:$s$ resonance of lowest order~$r$, which
dominates the numerically and experimentally relevant regime where
$\gamma>10^{-15}$.

The area covered by the sub-regions of such a resonance can be very small, see
the inset of Fig.~\ref{fig:Results}(a).
Therefore, it is necessary to search for resonances systematically.
To this end we determine the frequencies of orbits within the regular region, as
described in Ref.~\cite{KulLoeMerBaeKet2014}.
We then identify the $r$:$s$ resonance of lowest order $r$, by searching for
the rational frequencies $2\pi s/r$ with smallest possible denominator.

Specifically, for the standard map parity implies that $r$ has to be an even
number in order to reflect the correct number of subregions forming the
resonance chain.
For the examples we consider in this paper we find a dominant
$10$:$3$ resonance for $\K=2.9$ and a dominant $6$:$2$ resonance
for both $\K=3.4$ and $\K=3.5$.

\subsection{Integrable Approximation of a Regular Region including a Resonance
Chain}
  \label{Sec:IntegrableApproximation}

In order to determine an integrable approximation of the regular region which
includes the dominant $r$:$s$ resonance, we use the method introduced in
Ref.~\cite{KulLoeMerBaeKet2014}.
Here, we briefly summarize the key points.\

The integrable approximations $H_{r:s}(q, p)$ of
Ref.~\cite{KulLoeMerBaeKet2014} is generated in two steps.
First the normal-form Hamiltonian is defined as
\begin{subequations} \label{eq:HrsActAng-all}
  \begin{align}
    \label{eq:HrsActAng}
\hspace*{-0.4cm}    \mathcal{H}_{r:s}(\theta, I) &= \mathcal{H}_{0}(I) +
  2\Vrs \left(\frac{I}{I_{r:s}}\right)^{r/2}\!\!\!\cos(r\theta + \phi_0),\\
  \label{eq:H0}
 \mathcal{H}_{0}(I) &= \frac{(I-I_{r:s})^{2}}{2 M_{r:s}} +
\sum_{n=3}^{N_{\text{disp}}} h_{n}(I-I_{r:s})^{n}.
  \end{align}
\end{subequations}
It contains the essential information on the regular region in the co-rotating
frame of the resonance.
This Hamiltonian is precisely the effective pendulum Hamiltonian used in
Ref.~\cite{LoeBaeKetSch2010, SchMouUll2011}.
Here, $\mathcal{H}_{0}(I)$ is a low order polynomial, chosen such that its
derivative fits the actions and frequencies of the regular region in the
co-rotating frame of the resonance.
The action of the resonant torus is $I_{r:s}$.
The parameters $M_{r:s}$ and $V_{r:s}$ are determined from the size of the
resonance regions in the mixed system as well as the stability of its central
orbit \cite{EltSch2005}.
Finally, $\phi_0$ is used to control the fix-point locations of the resonance
chain.

In a second step, a canonical transformation
\begin{align}
  \label{eq:CanTrans}
  \CanTrans: (\theta, I) \mapsto (q,p)
\end{align}
is used to adapt the tori of the effective pendulum Hamiltonian to the shape of
the regular region in $(q,p)$-space, giving the Hamilton function
\begin{align}
  \label{eq:Hrs}
  H_{r:s}(q,p) = \mathcal{H}_{r:s}(\CanTrans^{-1}(q,p)).
\end{align}
The transformation $\CanTrans$ is composed of:\
(i) a harmonic oscillator transformation to the fixed point of the regular
region $\CanTrans^{0}$, Eq.~\eqref{eq:CanTransInit}, which provides a rough
integrable approximation and
(ii) a series of canonical near-identity transformations $\CanTrans^{1}, ...,
\CanTrans^{N_{\CanTrans}}$, Eq.~\eqref{eq:CanTransIter}, which improve the
agreement between the shape of tori of the mixed system and the integrable
approximation.

Note that a successful prediction of decay rates requires an integrable
approximation which provides a smooth extrapolation of tori into the chaotic
region \cite{BaeKetLoeSch2008, BaeKetLoe2010}, see insets of
Fig.~\ref{fig:Results}.
This is ensured by using simple near-identity transformations
$\CanTrans^{1}, ..., \CanTrans^{N_{\CanTrans}}$, i.\,e., low orders $N_{q},
N_{p}$ in Eq.~\eqref{eq:CanTransIter}.
For further details the reader is referred to Ref.~\cite{KulLoeMerBaeKet2014}
and Appendix~\ref{App:IntegrableApproximation}, where it is described how the
integrable Hamiltonians for the standard map at $\K=2.9$, $\K=3.4$ and $\K=3.5$,
see insets of Fig.~\ref{fig:Results}, are generated.

\subsection{Quantization of the Integrable Approximation}
  \label{Sec:Quantization}

In the following, we summarize the quantization procedure for the integrable
approximation.
The details are discussed in App.~\ref{App:IntegrableApproximationQuantum}.
In its final form, this quantization procedure is almost identical to the
approach presented in Ref.~\cite{LoeBaeKetSch2010}.
It consists of two steps:\
(Q1) The integrable approximation without resonance is used to construct states
which localize along a single quantizing torus of the regular region.
(Q2) The mixing of states, localizing along a single quantizing torus, is
described by solving the quantization of the effective pendulum Hamiltonian,
Eq.~\eqref{eq:HrsActAng-all}, introduced in Ref.~\cite{SchMouUll2011}.
Combining (Q1) and (Q2) gives the sought-after eigenstate $\ket{\mint}$ of the
integrable approximation which includes the resonance.

\begin{figure}[tb!]
  \begin{center}
\includegraphics[]{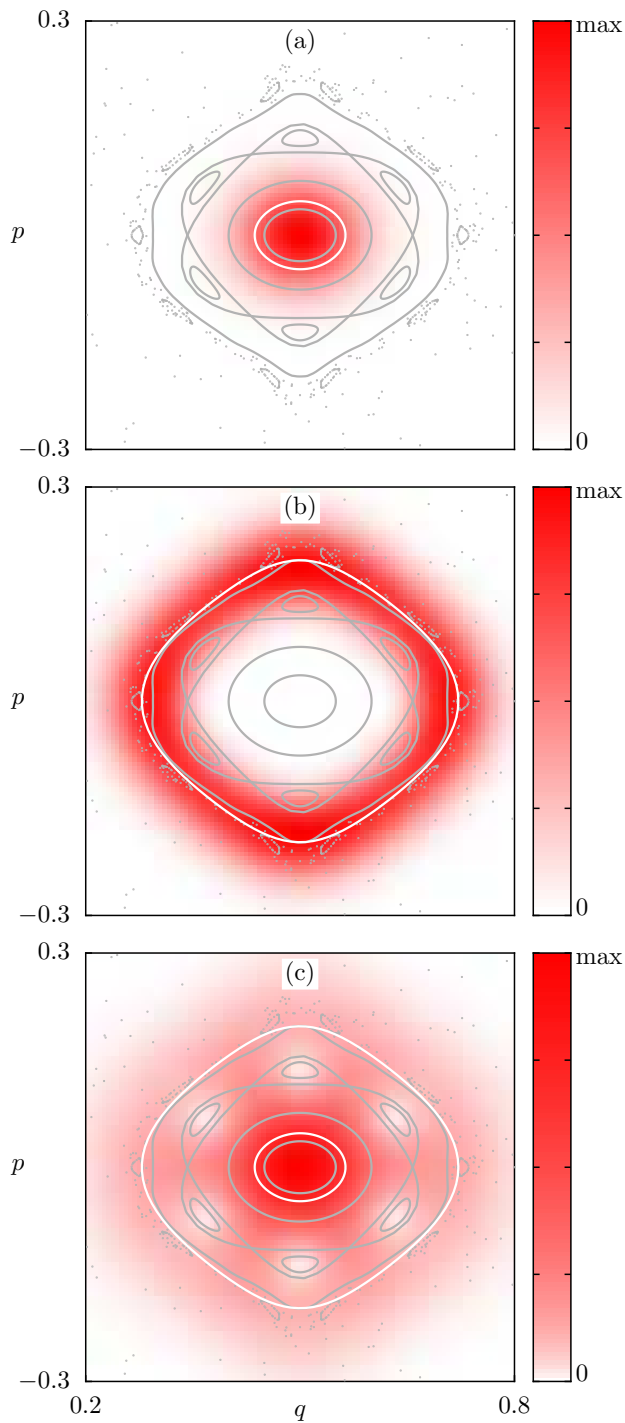}
\caption{(color online)
(a, b) Husimi representation of $\ket{I_n}$ for the standard map at $\K=3.4$
for (a) $n=0$ and (b) $n=6$ at $1/\heff=53$. Regular tori (gray lines) and
chaotic orbits (dots) illustrate the phase space. The quantizing tori of
$H_{r:s}$ for $V_{r:s}=0$ are shown by a thick (white) line.
(c) Approximate mode $\ket{\mint}$ for $m=0$.}
     \label{fig:Istates}
  \end{center}
\end{figure}

More specifically:\
(Q1) We use the canonical transformation, Eq.~\eqref{eq:CanTrans}, in order to
define the function $I(q,p)$.
Its contours approximate the tori of the regular phase-space region, ignoring
the resonance chain.
It thus resembles the role of the integrable approximation, previously used in
Refs.~\cite{BaeKetLoeSch2008, BaeKetLoe2010, LoeBaeKetSch2010}.
The Weyl-quantization of this function on a phase-space torus gives a Hermitian
matrix
\begin{align}
    \label{eq:WeylI}
  &\braOpket{\overline{q}_n}{\hat{I}}{\overline{q}_m} = \frac{1}{2N}
\sum_{l=0}^{2N-1}
\exp\left(\frac{i}{\hbar}(\overline{q}_n\!-\overline{q}_m)\,\overline{p}_{\frac{
l } {2 }}
\right)\times \\
&\quad \quad\left[I\!\left(\frac{\overline{q}_n\!+\overline{q}_m}{2},
\overline{p}_{\frac{l}{2}}\right)  + (-1)^l
I\!\left(\frac{\overline{q}_n \! +\overline{q}_m \!+ M_q}{2},
\overline{p}_{\frac{l}{2}}\right)\right]\!. \nonumber
\end{align}
Solving its eigenvalue equation gives states $\braket{\overline{q}_l}{I_n}$
which localize along a single contour of $I(q,p)$ with quantizing action $I_n
= \hbar(n+1/2)$.
These states model the localization of states along the tori of quantizing
action $\act_n$ in the mixed system.
For an illustration see Fig.~\ref{fig:Istates}(a,b).

(Q2) In the second step, we model the mixing of states
$\braket{\overline{q}_l}{I_n}$ due to the nonlinear resonance chain.
To this end, we follow Refs.~\cite{LoeBaeKetSch2010, SchMouUll2011} and
consider the quantization of the effective pendulum Hamiltonian,
Eq.~\eqref{eq:HrsActAng-all}, given by
\begin{align}
  \label{eq:HrsActAngIBasis}
\braOpket{I_m}{\widehat{\mathcal{H}}_{r:s}}{I_n} &=
\mathcal{H}_0(I_n)\,\delta_{m,n} +
V_{r:s} \left(\frac{\hbar}{\Irs}\right)^{r/2} \times \\
& \hspace*{-1.5cm} \left(e^{-\myi \phi_{0}}
\sqrt{\frac{n!}{(n-r)!}}\,\delta_{m,n-r} +
e^{\myi \phi_{0}}\sqrt{\frac{(n+r)!}{n!}}\,\delta_{m,n+r}\right). \nonumber
\end{align}
Solving this eigenvalue problem gives the sought-after state in the basis of
quantizing actions $\braket{I_n}{\mint}$.
Note, that the matrix in Eq.~\eqref{eq:HrsActAngIBasis} couples basis states
$\ket{I_n}$ and $\ket{I_{n'}}$ only if $|n'-n|=kr$.
Thus, the coefficients $\braket{I_{n}}{\mint}$ are non-zero, only if $n = m +
kr$.
This is called the selection rule of resonance-assisted tunneling.
Combining (Q1) and (Q2) results in the mode expansion
\begin{align}
  \label{eq:ModeExpansion}
  \braket{\overline{q}_l}{\mint} = \sum_{k}\braket{\overline{q}_l}{I_{m +
kr}} \braket{I_{m+kr}}{\mint}.
\end{align}
For an illustration of a state $\ket{\mint}$ see Fig.~\ref{fig:Istates}(c).
Note that its Husimi-function exhibits exactly the morphology discussed in 
Ref.~\cite{Wis2014}.

We now make a couple of remarks:\
(a) We use the above quantization procedure, rather than directly applying the
Weyl-rule to $H_{r:s}(q,p)$, Eq.~\eqref{eq:Hrs}, in order to explicitly enforce
the selection rule of resonance-assisted tunneling.
(b) The ad-hoc two step quantization scheme avoids the problem of defining the
quantum counterpart for the canonical transformations $\CanTrans^{1}, ...,
\CanTrans^{N_{\CanTrans}}$, Eq.~\eqref{eq:CanTransIter}, used in the classical
construction of the integrable approximation, see
App.~\ref{App:IntegrableApproximationQuantum} for details.
(c) The above quantization is almost identical to the procedure used in
Refs.~\cite{LoeBaeKetSch2010, SchMouUll2011}.
This allows for a direct comparison to the results of
Refs.~\cite{LoeBaeKetSch2010, SchMouUll2011}, see
Sec.~\ref{Sec:ResultsPerturbation}.
(d) The quantization procedure cannot determine the relative phase between
the terms in the mode expansion of Eq.~\eqref{eq:ModeExpansion}.

In order to understand the relative phase recall:\
(i) The coefficient vector $\braket{I_{m+kr}}{\mint}$ is determined by solving
the eigenvalue problem of Eq.~\eqref{eq:HrsActAngIBasis}.
Hence, it is determined up to a global phase $\xi_{m}$.
(ii) The coefficient vectors $\braket{\overline{q}_l}{I_{m+kr}}$ are determined
by solving the eigenvalue problem of Eq.~\eqref{eq:WeylI}.
Hence, each coefficient vector is determined up to a global phase
$\varphi_{m+kr}$.
Therefore:
(i) Changing the phase of the coefficient vector $\braket{I_{m+kr}}{\mint}$
in Eq.~\eqref{eq:ModeExpansion} changes the global phase of
$\braket{\overline{q}_l}{\mint}$.
This has no consequences for predicting decay rates.
However, (ii) changing the phases $\varphi_{m+kr}$ of each coefficient vector
$\braket{\overline{q}_l}{I_{m+kr}}$, changes the relative phase of
contributions in Eq.~\eqref{eq:ModeExpansion}.
This changes the interference between the contributions to the sum in
Eq.~\eqref{eq:ModeExpansion} and affects the predicted decay rates.

So far the phase issue was avoided by neglecting interference terms in the
tunneling predictions \cite{LoeBaeKetSch2010, SchMouUll2011}.
For the symmetrized standard map, we propose to define the phases as follows:\
(i) Eq.~\eqref{eq:HrsActAngIBasis} gives a real symmetric matrix.
This allows for choosing real coefficients $\braket{I_{n}}{\mint}$ such that
$\braket{I_{m}}{\mint}>0$.
(ii) Eq.~\eqref{eq:WeylI} also gives a real symmetric matrix.
This allows for choosing real coefficients $\braket{\overline{q}_l}{I_{n}}$.
Choosing the sign of these coefficients is discussed in
App.~\ref{App:IntegrableApproximationQuantum}.
The main idea is to exploit the eigenstates of the harmonic oscillator which
approximates the central fixed point of the regular region.
For these harmonic oscillator states the relative phase is well-defined.
Then we choose the sign of $\braket{\overline{q}_l}{I_{n}}$ such that its
overlap with the corresponding eigenstate of the harmonic oscillator is
positive, Eq.~\eqref{eq:ModeAligning}.

\section{Results}
  \label{Sec:Results}

\begin{figure}[tb]
\begin{center}
\includegraphics[]{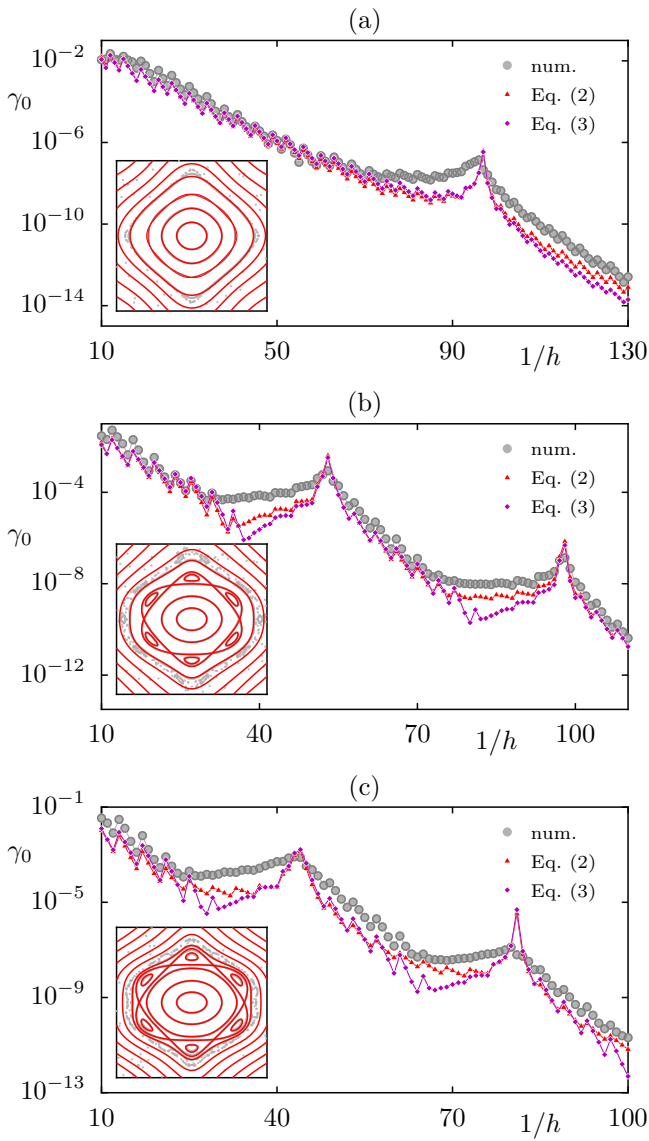}
  \caption{(color online) Decay rates for the standard map at (a) $\K=2.9$,
(b) $\K = 3.4$, and (c) $\K=3.5$ versus the inverse effective Planck constant
$1/h$. Numerically determined rates (dots) are compared to predicted rates,
using Eq.~\eqref{eq:GammaPrediction} ([red] triangles) and
Eq.~\eqref{eq:GammaPredictionAlt} ([magenta] squares). The insets show the
corresponding phase space with regular tori ([gray] lines) and chaotic orbits
(dots) with tori of the integrable approximation ([red] lines).}
     \label{fig:Results}
  \end{center}
\end{figure}

We now apply the above procedure to the standard map at $\K=2.9$, $3.4$, and
$3.5$.
This gives eigenstates $\ket{\mint}$ which we insert into our predictions,
Eqs.~\eqref{eq:GammaPrediction} and \eqref{eq:GammaPredictionAlt}.
The necessary time-evolution operator, used in Eq.~\eqref{eq:GammaPrediction},
is given by Eq.~\eqref{eq:StandardQmapOnTorus}.
The projector is defined by Eq.~\eqref{eq:Pabs} using $q_l=0.27, 0.26, 0.25$,
respectively.
The results are shown in Fig.~\ref{fig:Results}.
The numerically determined rates and the
predicted rates are overall in good qualitative agreement.
In both cases they deviate from the exact numerical rates by at most two orders
of magnitude.
In that the accuracy of the perturbation-free predictions,
Eq.~\eqref{eq:GammaPrediction} and Eq.~\eqref{eq:GammaPredictionAlt}, is
equivalent to perturbative predictions from Refs.~\cite{LoeBaeKetSch2010,
SchMouUll2011}.
This establishes Eqs.~\eqref{eq:GammaPrediction} and
\eqref{eq:GammaPredictionAlt} as state of the art perturbation-free predictions
of resonance-assisted regular-to-chaotic tunneling.
See Sec.~\ref{Sec:ResultsPerturbation} for a detailed comparison.

\subsection{Incoherent Predictions and Quantum Phase}
  \label{Sec:ResultsIncoherent}

As discussed in Sec.~\ref{Sec:Quantization} our quantization scheme cannot
determine the relative phases between the contributions of
Eq.~\eqref{eq:ModeExpansion} for a system without time-reversal symmetry.
In the following, we discuss the consequences of such an undetermined phase for
the prediction of decay rates.
To this end we summarize our predictions, Eqs.~\eqref{eq:GammaPrediction} and
\eqref{eq:GammaPredictionAlt}, in the following compact form
\begin{align}
  \label{eq:GammaPredictionCompact}
  \Gamma_{m}(t) := \big\|\pabs \qmap^{t} \ket{\mint} \big\|^2,
\end{align}
where $t=1$ denotes the prediction based on time-evolution and $t=0$ denotes
the prediction without time evolution.
Now we insert the mode expansion, Eq.~\eqref{eq:ModeExpansion}, and average
over the undetermined phases $\varphi_{m+kr}$ of the coefficient vectors
$\braket{\overline{q}_l}{I_{m+kr}}$.
This gives the incoherent prediction
\begin{align}
  \label{eq:GammaPredictionCompactIncoherent}
\Gamma_{m}^{\inc}(t) := \sum_{k} \Gamma_{m,m+kr}^{\diag}(t)
\end{align}
where the diagonal term $\Gamma_{m,n}^{\diag}(t)$ is the contribution of the
state $\ket{I_n}$ to the incoherent prediction as
\begin{align}
    \label{eq:GammaPredictionDiagonal}
 \Gamma_{m,n}^{\diag}(t) := \left|\braket{I_{n}}{\mint}\right|^{2}
\Gamma_{n}^{\text{d}}(t)
\end{align}
and
\begin{align}
  \label{eq:GammaPredictionCompactDirekt}
  \Gamma_{n}^{\text{d}}(t) := \big\|\pabs \qmap^{t} \ket{I_n} \big\|^2.
\end{align}
is the rate of direct regular-to-chaotic tunneling as previously introduced in
Refs.~\cite{BaeKetLoeSch2008, BaeKetLoe2010}.

\begin{figure}[tb]
\begin{center}
\includegraphics[]{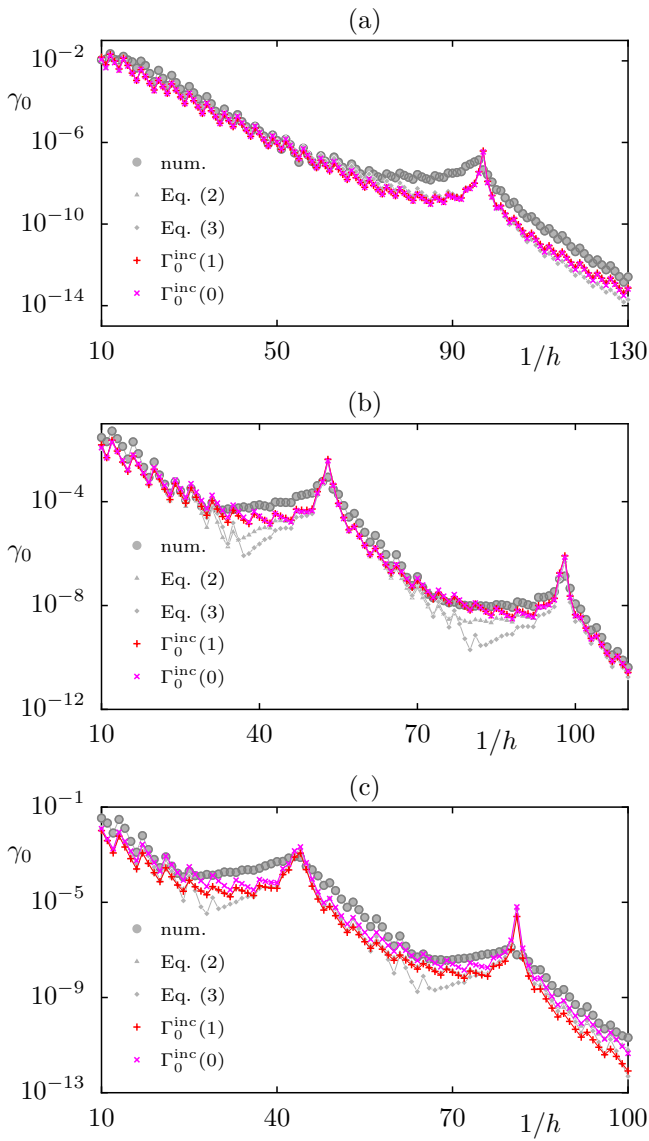}
  \caption{(color online) Decay rates for the standard map at (a) $\K=2.9$,
(b) $\K = 3.4$, and $\K=3.5$ versus the inverse effective Planck constant
$1/h$.
Numerically determined rates (dots) are compared to rates, predicted from
incoherent terms according to Eq.~\eqref{eq:GammaPredictionCompactIncoherent},
with $\Gamma_{m}^{\inc}(1)$ ([red] pluses) and $\Gamma_{m}^{\inc}(0)$ ([magenta]
crosses).
We further show predictions according to Eq.~\eqref{eq:GammaPrediction}
([gray] triangles) and Eq.~\eqref{eq:GammaPredictionAlt} ([gray] squares).
     \label{fig:ResultsIncoherent}}
  \end{center}
\end{figure}

The results based on Eq.~\eqref{eq:GammaPredictionCompactIncoherent} are
shown in Fig.~\ref{fig:ResultsIncoherent}.
As expected the incoherent predictions,
Eq.~\eqref{eq:GammaPredictionCompactIncoherent}, and the full predictions,
Eq.~\eqref{eq:GammaPredictionCompact}, agree very well in the regime where a
single diagonal contribution dominates, i.\,e., in the regime of direct
tunneling as well as the peak region.
However, in between these regions there are always two diagonal contributions of
similar magnitude, which can interfere.
It is in these regions that we observe clear deviations between the predictions
of Eq.~\eqref{eq:GammaPredictionCompact} and the incoherent predictions of
Eq.~\eqref{eq:GammaPredictionCompactIncoherent}.
In particular, for $\K=3.4$ and $\K=3.5$ the prediction of
Eq.~\eqref{eq:GammaPredictionCompact} predicts destructive interference, while
the incoherent results describe the numerical rates much better.

These results highlight the relevance of the phase factor $\varphi_{m+kr}$ for
obtaining an accurate description of decay rates even between the 
resonance-assisted tunneling peaks.
In previous studies of resonance-assisted tunneling in systems with a mixed 
phase space \cite{LoeBaeKetSch2010} this phase factor has been ignored by 
directly employing the incoherent predictions.
Hence, a satisfactory theoretical treatment of the phase factor $\varphi_{m+kr}$
does so far not exist.
Clearly, our current approach is also insufficient.
The precise reason is not clear to us.
We expect that exploiting the symmetry of the integrable approximation in order
to find a real representation of the approximate mode
$\braket{\overline{q}_l}{\mint}$ is too naive.
In particular, because it is used for approximating the metastable state
$\braket{\overline{q}_l}{m}$ of the open standard map, which can never admit an
entirely real representation.
For a detailed discussion of this point see  Sec.~\ref{Sec:ErrorAnalysis}.
Another possibility is that the phase factor in a non-integrable system is
beyond an integrable approximation.

\subsection{Perturbative Predictions}
  \label{Sec:ResultsPerturbation}

\begin{figure}[tb]
\begin{center}
\includegraphics[]{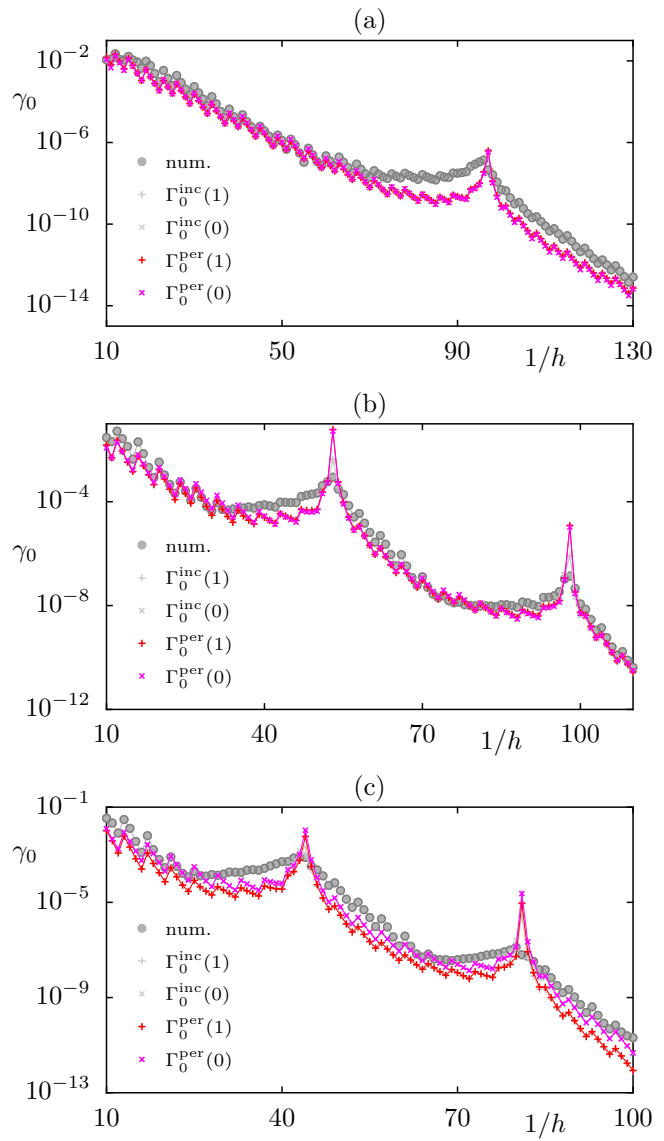}
  \caption{(color online) Decay rates for the standard map at (a) $\K=2.9$,
(b) $\K = 3.4$, and (c) $\K=3.5$ versus the inverse effective Planck constant
$1/h$.
Numerically determined rates (dots) are compared to rates, predicted
perturbatively according to Eq.~\eqref{eq:GammaPredictionCompactPerturbation}
with $\Gamma_{m}^{\per}(1)$ ([red] pluses) and $\Gamma_{m}^{\per}(0)$ ([magenta]
crosses).
We further show the prediction based on incoherent terms, according to
Eq.~\eqref{eq:GammaPredictionCompactIncoherent} with $\Gamma_{m}^{\inc}(1)$
([gray] pluses) and $\Gamma_{m}^{\inc}(0)$ ([gray] crosses).}
     \label{fig:ResultsPerturbation}
  \end{center}
\end{figure}

In this section, we compare our results to the perturbative
predictions of Refs.~\cite{LoeBaeKetSch2010, SchMouUll2011}.
This perturbative prediction is obtained by approximating the coefficient
$\braket{I_{m+kr}}{\mint}$ in the incoherent prediction
Eq.~\eqref{eq:GammaPredictionCompactIncoherent} by solving
Eq.~\eqref{eq:HrsActAngIBasis} perturbatively, \cite{SchMouUll2011},
\begin{align}
  \label{eq:PerturbationCoefficients}
\braket{I_{m+kr}}{\mint} \approx \mathcal{A}_{m,m+kr}^{(r:s)} := \prod_{l=1}^{k}
\frac{\braOpket{I_{m+lr}}{\widehat{\mathcal{H}}_{r:s}}{I_{m+(l-1)r}}}{\mathcal{H
} _ { 0 } (I_m) - \mathcal{H}_{0}(I_{m+kr})},
\end{align}
Note that $\mathcal{H}_{0}(I)$ is considered in the co-rotating frame.
This leads to
\begin{align}
  \label{eq:GammaPredictionCompactPerturbation}
\Gamma_{m}^{\per}(t) := \sum_{k} \left|\mathcal{A}_{m,m+kr}^{(r:s)}\right|^{2}
\Gamma_{m}^{\text{d}}(t).
\end{align}
A slight difference of the above expression as compared to
Ref.~\cite{LoeBaeKetSch2010, SchMouUll2011} is the use of the projector $\pabs$
rather than a projector on the whole chaotic region.
Thus our prediction eliminates a free parameter from the perturbative
predictions of Refs.~\cite{LoeBaeKetSch2010, SchMouUll2011}.
The results of the perturbative predictions are presented in
Fig.~\ref{fig:ResultsPerturbation}.
They agree with the prediction obtained from
Eq.~\eqref{eq:GammaPredictionCompactIncoherent}, with the slight difference
that the perturbative results deviate around the peak region.

We conclude this section with a short list of advantages and disadvantages of
the perturbation-free and perturbative predictions:\

(i) The perturbation-free framework, Eqs.~\eqref{eq:GammaPrediction} and
\eqref{eq:GammaPredictionAlt}, as well as their incoherent version,
Eq.~\eqref{eq:GammaPredictionCompactIncoherent}, predict numerical rates with
similar accuracy as the perturbative framework of Refs.~\cite{LoeBaeKetSch2010,
SchMouUll2011}.

(ii) One advantage of the perturbative prediction is the possibility to evaluate
the terms $\braket{I_{m}}{\mint}$ analytically,
Eq.~\eqref{eq:PerturbationCoefficients}.
Yet, for practical use even the perturbative approach requires an integrable
approximation for predicting the direct rates $\Gamma_{m}^{\text{d}}$.
Hence, both predictions are equally challenging in their implementation.

(iii) Another advantage of the perturbative prediction is the possibility to
include multiple resonances into
Eq.~\eqref{eq:GammaPredictionCompactPerturbation}, which is not yet
possible for the perturbation-free predictions presented in this paper.
Note that this restriction is not too severe, because decay rates in the
experimentally and numerically accessible regimes ($\gamma>10^{-15}$) are
typically affected by a single resonance only.
Nevertheless, an extension of the perturbation-free results to the
multi-resonance regime is of theoretical interest and requires normal-form
Hamiltonians $\mathcal{H}_{r:s}$ which include multiple resonances.

(iv) The main advantage of the perturbation-free framework is that it provides
the foundation for deriving a future semiclassical prediction of
resonance-assisted regular-to-chaotic tunneling \cite{FriMerLoeBaeKet}.

\section{Discussion}
  \label{Sec:Discussion}

In this section, we discuss several aspects of our results in detail.
In Sec.~\ref{Sec:BeyondRTCTunneling} we discuss the dependence of decay rates on
the choice of the leaky region.
In Sec.~\ref{Sec:DiscussionStates} we compare the metastable states $\ket{m}$
and $\ket{m'}$ to the eigenstate $\ket{\mint}$ of an integrable approximation.
In Sec.~\ref{Sec:ErrorAnalysis} we analyze the approximation of $\ket{m}$ and
$\ket{m'}$ via $\ket{\mint}$ more systematically.
In Sec.~\ref{Sec:PeakPosition} we comment on the predictability of peaks.

\subsection{Dependence of Decay Rates on the Leaky Region}
  \label{Sec:BeyondRTCTunneling}

\begin{figure}[tb]
\begin{center}
\includegraphics[]{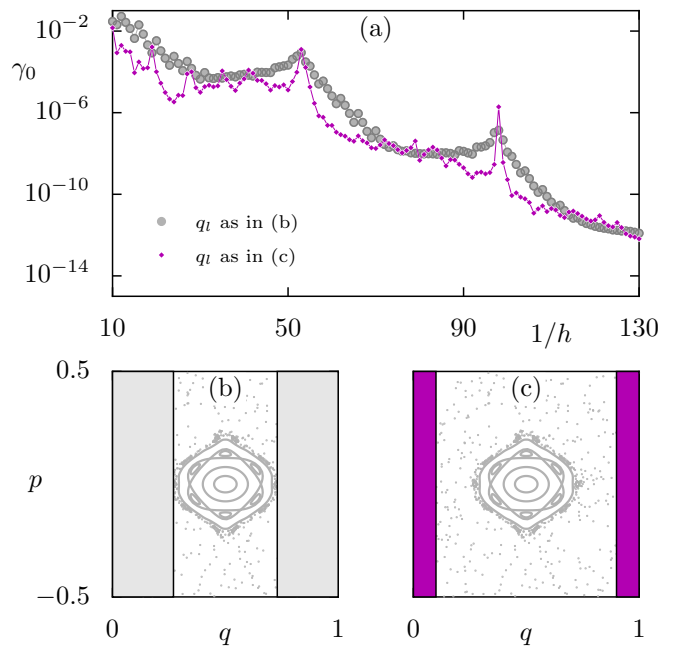}
  \caption{(color online) Numerically determined decay rates $\gamma_0$ of the
standard map at $\K=3.4$ versus the inverse effective Planck constant $1/\heff$
for $q_l=0.26$ ([gray] dots) and $q_l=0.1$ ([magenta] squares). (b, c) Phase
space with shaded areas showing the leaky regions corresponding to
$q_l$.
}
     \label{fig:DiscussionLeakyRegionK34}
  \end{center}
\end{figure}
\begin{figure}[tb]
\begin{center}
\includegraphics[]{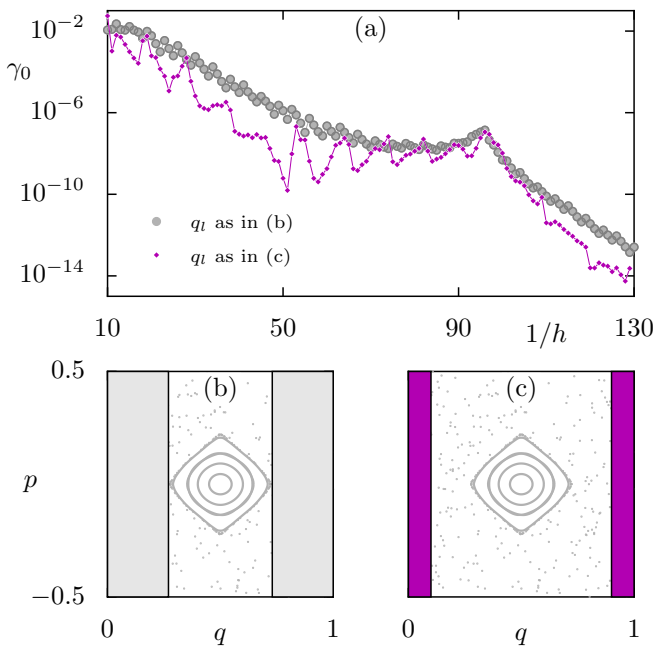}
  \caption{(color online) Same as Fig.~\ref{fig:DiscussionLeakyRegionK34} for
$\K=2.9$ with $q_l=0.27$ ([gray] dots) and $q_l=0.1$ ([magenta] squares)}
     \label{fig:DiscussionLeakyRegionK29}
  \end{center}
\end{figure}
This paper focuses entirely on situations where the leaky region $\mathcal{L}$
is chosen close to the regular--chaotic border region.
However, in generic Hamiltonian systems like the standard map, the chaotic
region is interspersed with partial barriers \cite{KayMeiPer1984a,
KayMeiPer1984b}.
This leads to sticky motion in a hierarchical region surrounding the regular
region.
Furthermore, the chaotic component might be inhomogeneous and exhibit slow
classical transport.

In view of these classical phenomena, it is not surprising that the numerical
decay rates of the standard map, defined via Eqs.~\eqref{eq:UopenEigenvalue},
depend on the choice of the leaky region via the parameter $q_l$.
In order to illustrate this phenomenon, we show the numerically determined decay
rate $\gamma_{0}$ of the standard map for two choices of the leaky region
and two different $\K$ parameters in Figs.~\ref{fig:DiscussionLeakyRegionK34}
and \ref{fig:DiscussionLeakyRegionK29}, respectively.

In Fig.~\ref{fig:DiscussionLeakyRegionK34} we show results for the standard map
at $\K=3.4$.
Here, we compare (i) the regular-to-chaotic decay rates obtained for
$q_l=0.26$ (parameter used in this paper, [gray] dots) to (ii) decay rates
obtained for $q_l=0.1$ ([magenta] squares).
While the decay rates for $q_l=0.26$ exhibit a rather smooth behavior the decay
rates for $q_l=0.1$ clearly exhibit additional oscillations and some overall
suppression.

An even stronger deviation between regular-to-chaotic decay rates with
varying leaky regions is observed in Fig.~\ref{fig:DiscussionLeakyRegionK29}
for the standard map at $\K=2.9$.
Here, (i) the decay rates as obtained for $q_l=0.27$ (parameter used in this
paper, [gray] dots) are compared to (ii) the decay rates obtained for $q_l=0.1$
([magenta] squares).
In addition to oscillations, the decay rates for $q_l=0.1$ exhibit a clear
suppression of their average value.

The origin of these deviations is unclear.
The suppression of decay rates for leaky regions far from the regular--chaotic
border could be due to slow transport through an inhomogeneous chaotic region
from the regular--chaotic border towards the leaky region.

So far a quantitative prediction of decay rates with leaky regions far from the
regular--chaotic border remains an open problem.
While varying the leaky region $\mathcal{L}$ close to the regular--chaotic
border can be accounted for by our approach, predicting decay rates with leaky
region far from the regular--chaotic border is beyond our framework.
In particular, while we observe that the numerical decay rates stabilize when
pushing the leaky region away from the regular--chaotic border, the predicted
rates continue to decrease exponentially.

So far the best approach for dealing with this problem is to use an effective
prediction \cite{LoeBaeKetSch2010, SchMouUll2011}.
To this end one argues that the numerical decay rate would not change much upon
pushing the boundary of the leaky region $\mathcal{L}$ beyond some effectively
enlarged regular region $\mathcal{R}_{\text{eff}}$.
See Ref.~\cite{SchMouUll2011} for a discussion of $\mathcal{R}_{\text{eff}}$.
Successively one would approximate the projector onto the leaky region
$\mathcal{L}$ in our predictions by the projector onto the complement of the
effectively enlarged regular region $\mathcal{L}_{\text{eff}}$.
This would result in an effective prediction $\Gamma_{m}^{\text{eff}}$.

Yet, there are several problems with such effective predictions:\
(a) Even though there are semiclassical arguments to define the effectively
enlarged regular region in terms of partial barriers \cite{SchMouUll2011},
replacing the leaky region $\mathcal{L}$ with some effective region
$\mathcal{L}_{\text{eff}}$ introduces an effective parameter to the prediction.
(b) Throughout this paper, we used leaky regions $\mathcal{L}$ which were
almost tangential to the effectively enlarged regular regions discussed in
Ref.~\cite{SchMouUll2011}.
Hence, replacing the region $\mathcal{L}$ with the effective region
$\mathcal{L}_{\text{eff}}$ would not give results which are too far away from
the predictions discussed in this paper, i.\,e., even the effective predictions
$\Gamma_{m}^{\text{eff}}$ clearly deviates from numerically determined decay
rates with leaky regions far from the regular region.
(c) Even when using $\mathcal{L}_{\text{eff}}$ as a free fit parameter the
effective prediction $\Gamma_{m}^{\text{eff}}$ can at most capture the average
behavior of numerical decay rates with leaky region far from the
regular--chaotic border.
In particular, the oscillations observed for the numerical rates in
Figs.~\ref{fig:DiscussionLeakyRegionK34} and \ref{fig:DiscussionLeakyRegionK29}
which span up to four orders of magnitude cannot be accounted for even by an
effective theory.

Note that, accurately predicting decay rates based on
Eqs.~\eqref{eq:GammaPrediction} and \eqref{eq:GammaPredictionAlt}, even for
leaky regions far from the regular region, requires modes $\ket{\mint}$ which
model the localization of $\ket{m}$ and $\ket{m'}$ even in the chaotic region.
We expect that this is beyond the framework of an integrable approximation.

\subsection{Metastable States and Integrable Eigenstates}
  \label{Sec:DiscussionStates}

\begin{figure}[tb]
\begin{center}
\includegraphics[]{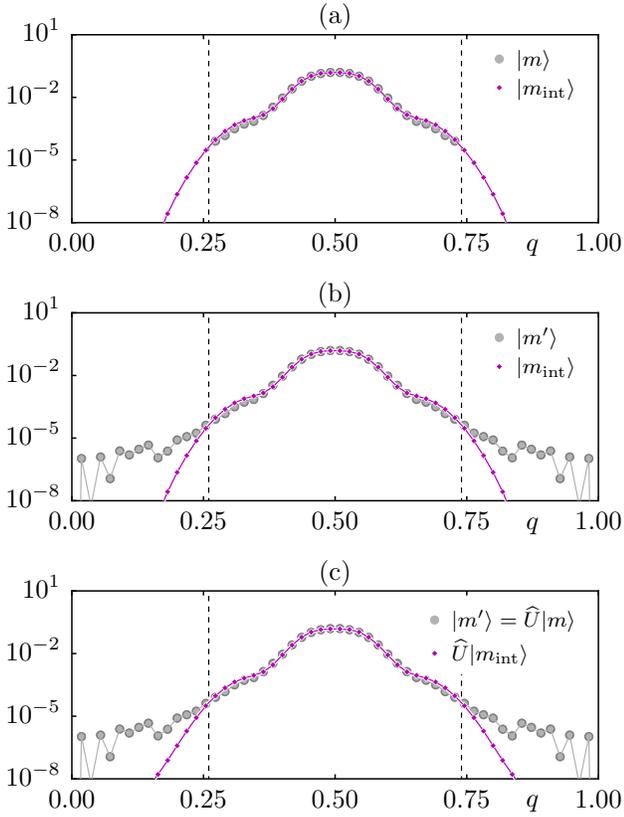}
  \caption{(color online) For the standard map at $\K=3.4$ with $1/\heff = 55$
we compare the position representation of
(a) $|\langle q|m\rangle|^2$ to $|\langle q|\mint\rangle|^2$,
(b) $|\langle q|m'\rangle|^2$ to $|\langle q|\mint\rangle|^2$, and
(c) $|\langle q|\qmap|m\rangle|^2$ to $|\langle q|\qmap|\mint\rangle|^2$,
depicting them by (gray) dots and (magenta) squares, respectively, for $m = m' =
\mint = 0$. The dashed lines mark the positions $q_l$ and $1-q_l$ of the leaky
region, as given in the text.}
     \label{fig:DiscussionStates}
  \end{center}
\end{figure}

We now discuss the key approximation of our predictions.
To this end we compare the metastable states $\ket{m}$ and $\ket{m'}$
to the corresponding approximate state $\ket{\mint}$,
which originates from an integrable approximation $H_{r:s}$ including
the relevant resonance.
We focus on a typical example using the states $m = m' = \mint = 0$ of the
standard map at $\K=3.4$ with $1/\heff=55$ close to the first resonance peak in
Fig.~\ref{fig:RAT_Intro}(c).
The absolute squared values of the states in position representation are
shown in Fig.~\ref{fig:DiscussionStates}.
Here we compare (a) $\ket{m}$ to $\ket{\mint}$, (b) $\ket{m'}$ to $\ket{\mint}$,
and (c) $\ket{m'} = \qmap\ket{m}$ to $\qmap\ket{\mint}$, depicting them by
(gray) dots and (magenta) squares, respectively.

As a first conclusion we see that the metastable states are well approximated by
their integrable partners within the non-leaky region, i.\,e., the region 
between the dashed lines in Figs.~\ref{fig:DiscussionStates}(a-c).
In particular, both the metastable states and their integrable approximations
exhibit the generic structure which is determined by the regular region and the
dominant $6$:$2$ resonance \cite{BroSchUll2001, BroSchUll2002}:\
(i) A main Gaussian-like hump at $q=0.5$ marks the main localization of the 
modes on the torus $\act_0$.
(ii) The decrease of the hump is interrupted at two side humps, which correspond
to the resonance-assisted contribution of each mode on the torus $\act_6$.
From there, the Gaussian-like exponential decrease continues towards the leaky 
region, which is outside the dashed lines in 
Figs.~\ref{fig:DiscussionStates}(a-c).

As a second conclusion from Fig.~\ref{fig:DiscussionStates} we infer that beyond
the regular--chaotic border, i.\,e., within the leaky region the metastable
states deviate from their integrable counter parts.
Here, the integrable states continue to decrease exponentially.
In contrast, the state $\ket{m}$ vanishes, see Fig.~\ref{fig:DiscussionStates}(a),
while the state $\ket{m'}=\qmap\ket{m}$, Eq.~\eqref{eq:TimeEvolvedMode},
does not decrease much slower, see Fig.~\ref{fig:DiscussionStates}(b,c).

Finally, we emphasize that $\ket{m'}$ and $\ket{\mint}$ agree for positions
close to the regular--chaotic border.
Furthermore, these contributions dominate the probability of $\ket{m'}$ and
$\ket{\mint}$ on the leaky region.
Precisely this ensures that replacing $\ket{m'}$ in the exact prediction,
Eq.~\eqref{eq:GammaPredOpenAlt}, by $\ket{\mint}$ results in a meaningful
prediction according to Eq.~\eqref{eq:GammaPredictionAlt}.
An analogous argument explains why replacing $\qmap\ket{m}$ in the exact result,
Eq.~\eqref{eq:GammaPredOpen}, by $\qmap\ket{\mint}$ gives meaningful predictions
according to Eq.~\eqref{eq:GammaPrediction}.

\subsection{Error Analysis}
  \label{Sec:ErrorAnalysis}

In this section, we investigate the approximation of the metastable states
$\ket{m}$ in the exact result Eq.~\eqref{eq:GammaPredOpen} via the mode
$\ket{\mint}$ in Eq.~\eqref{eq:GammaPrediction} from the perspective of
Eq.~\eqref{eq:ModeExpansion}, i.\,e.,
(i) we investigate the basis states $\ket{I_n}$ and (ii) the expansion
coefficients $\braket{I_n}{\mint}$.
We focus on the standard map at $\kappa=3.4$.

(i) In order to investigate our basis set $\ket{I_n}$, we expand the metastable
state $\ket{m}$ in this basis and insert this expansion into the exact
result~\eqref{eq:GammaPredOpen}.
This gives
\begin{align}
\label{eq:GammaPredOpenExpanded}
\gamma_{m} &= \sum_{n} \left|\braket{I_n}{m}\right|^{2} \big\|\pabs \qmap
\ket{I_n} \big\|^2 \\ \nonumber
&+ \sum_{n,n'} \braket{m}{I_{n'}}
\braOpket{I_{n'}}{\qmap^{\dagger}\pabs^2\qmap}{I_{n}}\braket{I_n}{m}.
\end{align}
Since the diagonal terms
\begin{align}
    \label{eq:GammaOpenDiag}
 \gamma_{m,n}^{\diag} := \left|\braket{I_n}{m}\right|^{2}\big\|\pabs \qmap
\ket{I_n} \big\|^2  = \left|\braket{I_n}{m}\right|^{2}
\Gamma_{n}^{\text{d}}(1)\end{align}
provide a bound to the off-diagonal terms according to Cauchy's inequality
\begin{align}
 \left|\braket{m}{I_{n'}}
\braOpket{I_{n'}}{\qmap^{\dagger}\pabs^2\qmap}{I_{n}}\braket{I_n}{m}\right| \le
\sqrt{\gamma_{m,n}^{\diag} \gamma_{m,n'}^{\diag}}
\end{align}
we can interpret them as a way to quantify the contribution of the $n$th basis
state $\ket{I_n}$ to the decay rate $\gamma_{m}$.
In that $\gamma_{m,n}^{\diag}$ takes a similar role as the contribution
spectrum,
discussed in Ref.~\cite{HanShuIke2015}.

\begin{figure}[tb!]
\begin{center}
\includegraphics[]{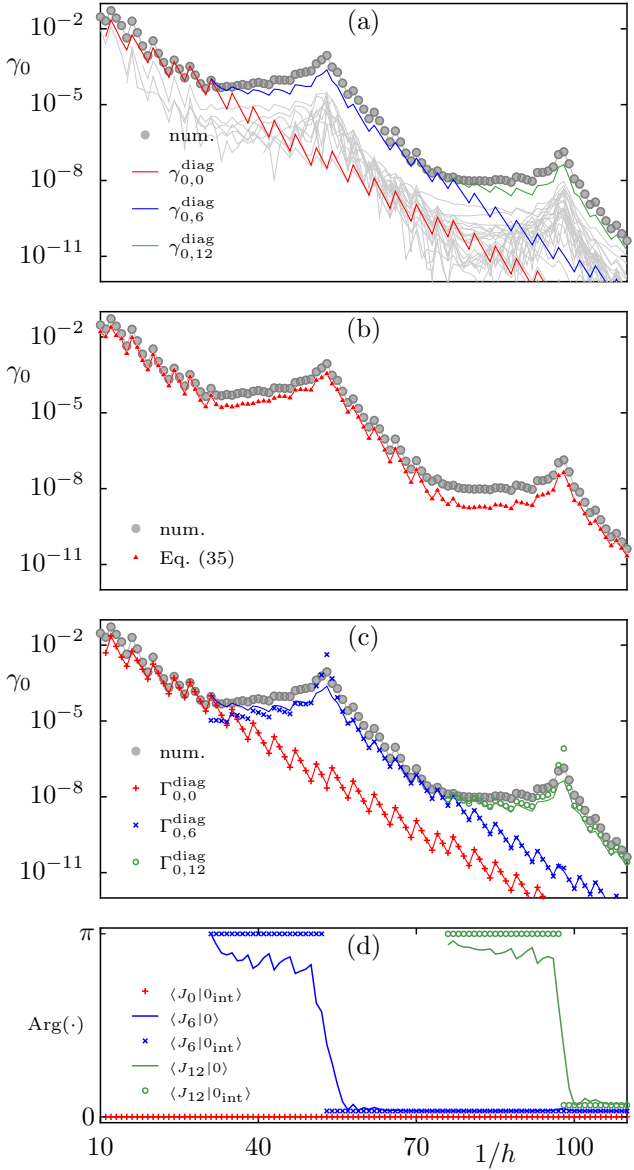}
  \caption{(color online) Error analysis for the standard map at $\K=3.4$.
(a,b,c) The numerically determined rates ([gray] dots) and (d) the numerically
determined phases $\Arg(\braket{I_n}{0})$ for $n=0,1,2$ (lines) are shown
versus the inverse effective Planck constant $1/\heff$.
(a) The contributions $\gamma_{0,n}^{\diag}$, Eq.~\eqref{eq:GammaOpenDiag}, are
shown by lines.
(b) The reduced prediction, Eq.~\eqref{eq:GammaPredOpenExpanded} with
$n,n'\in\{0,6,12\}$ is shown by (red) triangles.
(c) The contributions $\gamma_{0,n}^{\diag}$ of Eq.~\eqref{eq:GammaOpenDiag}
(lines), are compared to $\Gamma_{0,n}^{\diag}(1)$ of
Eq.~\eqref{eq:GammaPredictionDiagonal} (markers) for $n=0,6,12$.
(d) The phases $\Arg(\braket{I_n}{0})$ (lines) and
$\Arg(\braket{I_n}{0_{\text{int}}})$ (markers) are compared for $n=0,6,12$.
[Phases close to zero are slightly shifted for $n=6,12$ for visibility.]
}
     \label{fig:ErrorAnalysis}
  \end{center}
\end{figure}
In Fig.~\ref{fig:ErrorAnalysis}(a), we consider all contributions
$\gamma_{0,n}^{\diag}$ (lines) in comparison with the
decay rate $\gamma_{0}$ (dots) for the standard map at $\K=3.4$.
While most contributions are two to three orders of magnitude smaller than
$\gamma_{0}$, we find that the contributions $\gamma_{0,0}^{\diag}$,
$\gamma_{0,6}^{\diag}$, and $\gamma_{0,12}^{\diag}$ dominate.
In order to further test whether the modes $\ket{I_{n}}$ with $n=0,6,12$ are
sufficient for describing $\gamma_{0}$ we sum the contributions $n,n'\in
\{0,6,12\}$ of the dominant terms in Eq.~\eqref{eq:GammaPredOpenExpanded}.
This gives the red curve of Fig.~\ref{fig:ErrorAnalysis}(b).
From this numerical observations we conclude that a reasonable description of
$\gamma_{0}$ can be extracted using an approximate mode exclusively
composed of states $\ket{I_n}$ with $n=0,6,12,...$, as used in
Eq.~\eqref{eq:ModeExpansion}.
However, it should be noted that the difference between $\gamma_{0}$ and its
reduced version, based on contributions $n,n'\in \{0,6,12\}$ in
Eq.~\eqref{eq:GammaPredOpenExpanded}, is already of the order of $\gamma_{0}$
itself.
See the region $70<1/\heff<100$ of Fig.~\ref{fig:ErrorAnalysis}(b) in
particular.
Hence, reducing the metastable state $\ket{m}$ to an approximate mode
$\ket{\mint}$ using only basis states $\ket{I_{m+kr}}$ as in
Eq.~\eqref{eq:ModeExpansion} can at best provide a reasonable backbone for
describing the structure of $\gamma_{0}$.
On the other hand, for our example a prediction of $\gamma_{0}$ where the
remainder is smaller than the decay rate based on a reduced set of basis states
$\ket{I_n}$ is only possible when summing over many additional contributions,
even including $n\neq m+kr$.
The precise origin of such contributions $\gamma_{m,n}^{\diag}$ with
$n\neq m+kr$ is currently under debate \cite{HanShuIke2015}:\
From the framework of resonance-assisted tunneling~\cite{BroSchUll2001,
BroSchUll2002, LoeBaeKetSch2010, SchMouUll2011}, we expect that the overlap
$\braket{I_n}{m}$ vanishes for $n\neq m+kr$.
Hence, one might argue that the contributions $\gamma_{m,n}^{\diag}$ with $n\neq
m+kr$ arise in our example only because our basis $\ket{I_n}$ is insufficiently
accurate to decompose $\ket{m}$ according to the theoretical expectation of
resonance-assisted tunneling.
On the other hand, the authors of of Ref.~\cite{HanShuIke2015} observe
non-vanishing contributions $\braket{I_n}{m}$ also for $n\neq m+kr$ even for a
near-integrable situation, where an excellent integrable approximations exist.
They argue that non-vanishing $\braket{I_n}{m}$ should always occur and claim
their treatment is beyond the current framework of resonance-assisted tunneling.
Independent of the origin of the non-zero contributions
$\gamma_{m,n}^{\diag}$ for $n\neq m+kr$, their theoretical description
is beyond the scope of this paper.
In our examples the irrelevance of these contributions is ensured by choosing
leaky regions close to the regular--chaotic border.
However, for leaky regions far from the regular--chaotic border the
contributions $\gamma_{m,n}^{\diag}$ with $n\neq m+kr$ become relevant.

(ii) In the next step we evaluate the errors introduced by replacing the
expansion coefficients $\braket{I_{m+kr}}{m}$ by $\braket{I_{m+kr}}{\mint}$ in
Eq.~\eqref{eq:GammaPredOpenExpanded}.
We focus on the corresponding diagonal contributions $\gamma_{m,m+kr}^{\diag}$
and $\Gamma_{m,m+kr}^{\diag}$, which represent the squared norm of the expansion
coefficients $\braket{I_{m+kr}}{m}$ and $\braket{I_{m+kr}}{\mint}$ up to a
multiplication by the direct rate $\Gamma_{m+kr}^{\text{d}}(1)$.
See lines and symbols in Fig.~\ref{fig:ErrorAnalysis}(c), respectively.
From this data we conclude that the norm of $\braket{I_{m+kr}}{\mint}$ provides
a reasonable approximations for the norm of the expansion coefficients
$\braket{I_{m+kr}}{m}$.
The deviations before each peak could be due to neglecting the higher order 
action dependencies discussed in Ref.~\cite{SchMouUll2011} in the Hamilton 
function of Eq.~\eqref{eq:HrsActAng-all}.
Furthermore, we expect that the slightly broader peaks in the numerical rates
$\gamma_{0,kr}^{\diag}$ as compared to the sharper peaks of
$\Gamma_{0,kr}^{\diag}(1)$ observed for the integrable approximation, are
related to the openness of the mixed system.

Finally, in Fig.~\ref{fig:ErrorAnalysis}(d) we compare the phases
$\Arg(\braket{I_{m+kr}}{m})$ and $\Arg(\braket{I_{m+kr}}{\mint})$, for $m=0$ and
$k=0,1,2$, respectively.
Here, $\Arg(\cdot)\in(-\pi,\pi]$ is the principal value of the complex argument
function.
Note that the global phase of $\ket{m}$ is fixed by setting
$\Arg(\braket{I_{m}}{m})=0$.
The phases $\Arg(\braket{I_{m+kr}}{\mint})$ are fixed as described in 
App.~\ref{App:IntegrableApproximationQuantum}.
While the phases of $\Arg(\braket{I_{m+kr}}{\mint})$ jump from $\pi$ to zero
upon traversing the peak for decreasing $1/\heff$ (change from destructive to
constructive interference) their counterparts for $\Arg(\braket{I_{m+kr}}{m})$
seem to follow this jump in a smoothed out way.
Compare symbols and lines in Fig.~\ref{fig:ErrorAnalysis}(d).

We attribute this phase detuning to the openness of the system, i.\,e.:\
(a) The symmetries of the integrable approximation allow for choosing a real
representation of the coefficient $\braket{I_{m+kr}}{\mint}$.
Its phase can thus only take values
$\Arg(\braket{I_{m+kr}}{\mint})\in\{0,\pi\}$.
In contrast (b) the mode $\ket{m}$ originates from an open system and thus the
coefficient $\braket{I_{m+kr}}{m}$ are usually complex such that
$\Arg(\braket{I_{m+kr}}{m})$ might take any value.

While the deviation between the numerically determined phases $\Arg( \braket{
I_{m+kr}}{m})$ and the theoretically predicted phases $\Arg(\braket{ I_{m+kr}}{
\mint})$ are seemingly small in Fig.~\ref{fig:ErrorAnalysis}(d), their deviation
has huge effects on the predicted decay rate, i.\,e.:\
(a) Eq.~\eqref{eq:GammaPrediction} predicts destructive interference of the
diagonal terms in the region before each peak.
This leads to strong deviations from the numerical decay rate, see
Fig.~\ref{fig:Results}(b).
On the other hand, (b) already the minimal detuning of
$\Arg(\braket{I_{m+kr}}{m})$ from our prediction
$\Arg(\braket{I_{m+kr}}{\mint})$ is sufficient to lift the destructive
interference.
We assume that this explains why the incoherent prediction,
Eq.~\eqref{eq:GammaPredictionCompactIncoherent}, as illustrated in
Fig.~\ref{fig:ResultsIncoherent}, describe the numerical rates
much better than predictions according to, Eq.~\eqref{eq:GammaPrediction}, see
Fig.~\ref{fig:ResultsIncoherent}.

\subsection{Predictability of Peak Positions}
  \label{Sec:PeakPosition}

\begin{figure}[tb!]
\begin{center}
\includegraphics[]{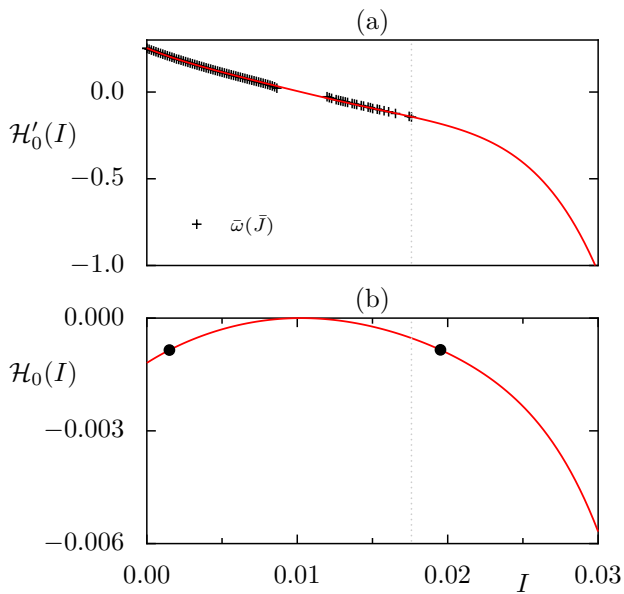}
  \caption{(color online) For the standard map at $\K=3.4$ we show (a) the fit
of $\mathcal{H}'_{0}(I)$ (line) to the actions and frequencies of the regular
region $(\bar{J},\bar{\omega})$ (crosses).
(b) The function $\mathcal{H}_{0}(I)$ is shown as a (red) line. The two (black)
dots show $(I_{m}, \mathcal{H}_{0}(I_{m}))$ and $(I_{m+kr},
\mathcal{H}_{0}(I_{m+kr}))$ at $1/\heff=53$.
(a, b) The dotted line shows the position of $\bar{J}_{\text{max}}$.
}
     \label{fig:dispersion}
  \end{center}
\end{figure}

Finally, we discuss the predictability of peak positions.
To this end we recall that $\mathcal{H}_{0}(I)$ in Eq.~\eqref{eq:H0}, is
determined by fitting its derivative to the numerically determined actions and
frequencies $(\bar{\omega}, \bar{\act})$ of the regular phase-space region in
the co-rotating frame.
For an illustration see Fig.~\ref{fig:dispersion}.
In particular, the data of the mixed system has a maximal action
$\bar{J}_{\text{max}}$, see (gray) dotted line in Fig.~\ref{fig:dispersion}.
Hence, $\mathcal{H}_{0}$ can be well controlled in the regular region
$I<\bar{J}_{\text{max}}$.
However, for $I>\bar{J}_{\text{max}}$ the function $\mathcal{H}_{0}$ is only an
extrapolation to the chaotic region.
Furthermore, the integrable approximation predicts a peak for $\gamma_{m}$
\cite{BroSchUll2001, BroSchUll2002, LoeBaeKetSch2010, SchMouUll2011}, if
\begin{align}
 \mathcal{H}_{0}(I_m) = \mathcal{H}_{0}(I_{m+kr}),
\end{align}
where $I_m=\hbar(m+1/2)$ and $I_{m+kr}=\hbar(m+kr+1/2)$.
This resonance conditions follows from Eq.~\eqref{eq:PerturbationCoefficients}.

However, for all examples presented in this paper the resonant torus
$I_{m+kr}$ is always located outside of the regular region, where
$\mathcal{H}_{0}(I)$ is only given by an extrapolation.
See Fig.~\ref{fig:Istates}(c) for an example of this situation.
The (black) dots in Fig.~\ref{fig:dispersion}(b) show the corresponding
situation for $\mathcal{H}_{0}(I)$.
In such a situation our approach cannot guarantee an accurate prediction of the
peak position.
Usually, this problem is not too severe and the extrapolation is good enough.
An example where this problem appears can be seen in the second peak of
Fig.~\ref{fig:Results}(a) where the peak of the numerical decay rates and the
predicted rates is shifted by $1/\heff=1$.

\section{Summary and Outlook}
  \label{Sec:Summary}

In this paper we present two perturbation-free predictions of resonance-assisted
regular-to-chaotic decay rates, Eqs.~\eqref{eq:GammaPrediction} and
\eqref{eq:GammaPredictionAlt}.
Both predictions are based on eigenstates $\ket{\mint}$ of an integrable
approximation $H_{r:s}$, Eq.~\eqref{eq:HrsEigenvalue}.
The key point is the use of an integrable approximation $H_{r:s}$ of the mixed
regular--chaotic system which includes the relevant nonlinear resonance chain.
Therefore $\ket{\mint}$ models the localization of regular modes on the regular
region, including resonance-assisted contributions in a non-perturbative way.
This allows for extending the validity of Eq.~\eqref{eq:GammaPrediction},
previously used for direct tunneling in Refs.~\cite{BaeKetLoeSch2008,
BaeKetLoe2010}, to the regime of resonance-assisted tunneling.
Furthermore, we introduce a second prediction,
Eq.~\eqref{eq:GammaPredictionAlt}, which no longer requires the time-evolution
operator.
Instead it allows for predicting decay rates using the localization of the
approximate mode on the leaky region.
In that Eq.~\eqref{eq:GammaPredictionAlt} provides an excellent foundation for a
future semiclassical prediction of resonance-assisted regular-to-chaotic
decay rates \cite{FriMerLoeBaeKet} in the spirit of Refs.~\cite{DeuMou2010,
DeuMouSch2013}.
The validity of the presented approach is verified for the standard map, where
predicted and numerically determined regular-to-chaotic decay rates show
good agreement.

Finally, we list future challenges:\
(a) The presented approach is so far limited to periodically driven systems
with one degree of freedom.
An extension to autonomous or periodically driven systems with two or more
degrees of freedom is an interesting open problem.
(b) The perturbation-free approach applies to the experimentally and
numerically relevant regime, where a single resonance dominates
regular-to-chaotic tunneling.
Its extension to the semiclassical regime where multiple-resonances affect
tunneling is of theoretical interest.
(c) The suppression of decay rates due to partial barriers is so far treated by
choosing leaky regions close to the regular--chaotic region.
Explicitly predicting the additional suppression of decay rates due to slow chaotic
transport through inhomogeneous chaotic regions remains an open question.

\begin{acknowledgments}

We gratefully acknowledge fruitful discussions with
J{\'e}r{\'e}my Le Deunff,
Felix Fritzsch,
Yasutaka Hanada,
Hiromitsu Harada,
Kensuke Ikeda,
Martin K{\"o}rber,
Steffen L\"ock,
Amaury Mouchet,
Peter Schlagheck,
and
Akira Shudo.
We acknowledge support by the Deutsche Forschungsgemeinschaft (DFG) Grant No.\
BA 1973/4-1.
N.M.\ acknowledges successive support by JSPS (Japan) Grant No.\ PE 14701 and
Deutsche Forschungsgemeinschaft (DFG) Grant No.\ ME 4587/1-1.
\end{acknowledgments}

\begin{appendix}

\section{Derivation of Eq.~\eqref{eq:GammaPredOpen}}
  \label{App:Derivation}

In this appendix we derive Eq.~\eqref{eq:GammaPredOpen} starting from
Eqs.~\eqref{eq:Uopen} and \eqref{eq:UopenEigenvalue}.
Taking the norm of the eigenvalue equation~\eqref{eq:UopenEigenvalue}
for a normalized state $\ket{m}$ one finds
\begin{align}
  \exp{\left(-\gamma_{m}\right)}
  &= \big\|\Uopen\ket{m}\big\|^{2}
  = \left\langle m \right| \Uopen^{\dagger}\Uopen \ket{m} \\
  &\hspace*{-1.7cm}
  = \left\langle m \right| (\oneOp-\pabs)^{\dagger}
\qmap^{\dagger}(\oneOp-\pabs)^{\dagger}(\oneOp-\pabs)\qmap
(\oneOp-\pabs)\ket{m}, \nonumber
\end{align}
where in the last step the definition of $\Uopen$, Eq.~\eqref{eq:Uopen}, is used.
We simplify this expression using
\begin{align}
  \label{eq:ProjectiveInvariance}
  (\oneOp-\pabs)\ket{m} = \ket{m},
\end{align}
which follows from Eqs.~\eqref{eq:Uopen} and \eqref{eq:UopenEigenvalue}, giving
\begin{align}
 \exp{\left(-\gamma_{m}\right)}
 &= \left\langle m \right|
\qmap^{\dagger}(\oneOp-\pabs)^{\dagger}(\oneOp-\pabs)\qmap \ket{m}.
\end{align}
Finally, exploiting the idempotence and hermiticity of the projector $\pabs$ gives
\begin{align}
 \exp{\left(-\gamma_{m}\right)}
 &= \left\langle m \right| \qmap^{\dagger}\qmap\ket{m}
  - \left\langle m \right| \qmap^{\dagger} \pabs \qmap \ket{m} \nonumber\\
 &= 1 - \|\pabs\qmap\ket{m}\|^{2},
\end{align}
where in the last step the unitarity of $\qmap$ is used. From
this follows the expression for regular-to-chaotic tunneling
rates, Eq.~\eqref{eq:GammaPredOpen}.

\section{Isospectrality}
  \label{App:Isospectrality}

In this appendix, we demonstrate the isospectrality of the sub-unitary operators
$\Uopen$ and $\UopenAlt$ as defined by Eqs.~\eqref{eq:Uopen} and
\eqref{eq:UopenAlt}, respectively.
Furthermore, we discuss the transformation relating their eigenmodes.
For convenience, we repeat the corresponding eigenvalue equations
\eqref{eq:UopenEigenvalue} and \eqref{eq:UopenAltEigenvalue}
\begin{align}
 \label{eq:UopenEigenvalueAppendix}
 \Uopen \ket{m} &= \lambda_m \ket{m}, \\
 \label{eq:UopenAltEigenvalueAppendix}
 \UopenAlt \ket{m'} &= \lambda_{m}' \ket{m'},
\end{align}
where the eigenvalues have been denoted by $\lambda_m$ and
$\lambda_m'$.

We now demonstrate the isospectrality of $\Uopen$ and $\UopenAlt$.
To this end we show:\

(a) For each eigenstate $\ket{m}$ of $\Uopen$ with eigenvalue
$\lambda_m$, $\qmap\ket{m}$ is an eigenstate of $\UopenAlt$ with the same eigenvalue
$\lambda_m$
\begin{align}
 \UopenAlt\qmap\ket{m}
&\stackrel{\eqref{eq:UopenAlt}}{=}\qmap(\oneOp-\pabs)\qmap
\ket{m} \\
&\stackrel{\eqref{eq:ProjectiveInvariance}}{=} \qmap(\oneOp-\pabs)\qmap
(\oneOp-\pabs)
\ket{m} \\
&\stackrel{\eqref{eq:Uopen}}{=} \qmap\Uopen\ket{m} \\
&\stackrel{\eqref{eq:UopenEigenvalueAppendix}}{=} \lambda_m \qmap
\ket{m}.
\end{align}
This further shows that the normalized eigenmode $\ket{m}$ of $\Uopen$ with
eigenvalue $\lambda_m$ gives a normalized eigenmode $\ket{m'}$ of $\UopenAlt$
with eigenvalue $\lambda_m$ according to Eq.~\eqref{eq:TimeEvolvedMode}.

(b) For each eigenstate $\ket{m'}$ of $\UopenAlt$ with eigenvalue
$\lambda_m'$, the state $(\oneOp-\pabs)\ket{m'}$ is an eigenstate of $\Uopen$
with
the same eigenvalue $\lambda_m'$
\begin{align}
 \Uopen (\oneOp-\pabs)\ket{m'}
&\stackrel{\eqref{eq:Uopen}}{=}(\oneOp-\pabs)\qmap
(\oneOp-\pabs)^2 \ket{m'} \\
&\;= (\oneOp-\pabs)\qmap (\oneOp-\pabs)\ket{m'} \\
&\stackrel{\eqref{eq:UopenAlt}}{=} (\oneOp-\pabs)\UopenAlt\ket{m'} \\
&\stackrel{\eqref{eq:UopenAltEigenvalueAppendix}}{=} \lambda_m' (\oneOp-\pabs)
\ket{m'}.
\end{align}
This further shows that for non-zero eigenvalue $\lambda_m'$ the normalized
eigenmode $\ket{m'}$ of $\UopenAlt$ gives a normalized eigenmode $\ket{m}$ of
$\Uopen$ according to Eq.~\eqref{eq:SameLocalization}.

\section{Details of the Integrable Approximation}
  \label{App:IntegrableApproximation}

In this appendix we summarize some technical aspects on the integrable
approximation.
Computational details of the classical integrable approximation as well as
slight changes as compared to Ref.~\cite{KulLoeMerBaeKet2014} are given in
Sec.~\ref{App:IntegrableApproximationClassical}.
Details of the quantization are discussed in
Sec.~\ref{App:IntegrableApproximationQuantum}.

\subsection{Details of the Classical Integrable Approximation}
  \label{App:IntegrableApproximationClassical}

We now summarize the modifications of the algorithm described in
\cite{KulLoeMerBaeKet2014} in order to account for the symmetries of our system.
Then we give a list of relevant computational parameters.

\subsubsection{Symmetrization}

In agreement with Ref.~\cite{KulLoeMerBaeKet2014} the canonical transformation
$\mathcal{T}$, Eq.~\eqref{eq:CanTrans}, is composed of (i) an initial canonical
transformation
\begin{align}
  \label{eq:CanTransInit}
  \CanTrans^{0}: (\theta, I) \mapsto (Q,P)
\end{align}
which provides a rough integrable approximation of the regular phase-space
region and (ii) a series of canonical near-identity transformations
\begin{align}
  \label{eq:CanTransIter}
  \CanTrans' \equiv \CanTrans^{N_{\CanTrans}}\circ...\circ\CanTrans^{1}: (Q, P)
\mapsto (q,p),
\end{align}
which improve the agreement between the shape of the tori of the mixed system
and the integrable approximation.

In contrast to Ref.~\cite{KulLoeMerBaeKet2014} we use the symmetrized standard
map in this paper.
In order to account for this symmetry, we specify the canonical transformation,
Eq.~\eqref{eq:CanTransInit}, as
\begin{align}
\label{eq:CanTransInitHO}
 \CanTrans^{0}:
 \begin{pmatrix}
    \theta \\ I
 \end{pmatrix}
 \mapsto
 \begin{pmatrix}
    Q \\ P
 \end{pmatrix}
 =
 \begin{pmatrix}
    q^{\star} + \sqrt{2I/\sigma}\cos(\theta) \\
    p^{\star} - \sqrt{2I\sigma}\sin(\theta)
 \end{pmatrix}
\end{align}
Here, $(q^{\star}, p^{\star}) = (0.5, 0.0)$ are the coordinates of the central
fixed point in the standard map.
The parameter $\sigma$ is determined from the stability matrix of the standard
map at $(q^{\star}, p^{\star})$
\begin{align}
\mathcal{M} =
 \begin{pmatrix}
  1-\frac{\K}{2} & 1 \\
  - \K (1 - \frac{\K}{4}) & 1-\frac{\K}{2}
 \end{pmatrix}
\end{align}
as \cite{LoeLoeBaeKet2013}
\begin{align}
 \sigma^{2} =
\frac{\left|1+\frac{\K}{2}\right|-\left|1-\frac{\K}{2}\right|}{
\left|1+\frac{\K}{2}\right|+\left|1-\frac{\K}{2}\right|}.
\end{align}

Furthermore, the symmetry of the systems is imposed on the transformations
$\CanTrans^{1}, ..., \CanTrans^{N_{\CanTrans}}$, Eq.~\eqref{eq:CanTransIter},
by specifying their generating function as
\begin{align}
  \label{eq:CanTransFamily}
    F^{a}(q,p') &=\\ \nonumber
    qp' + &\sum_{n=1}^{N_{q}}\sum_{m=1}^{N_{p}} a_{m,n}
\sin(2\pi n[q-q^{\star}]) \sin(2\pi m[p'-p^{\star}]),
\end{align}
rather than using the more general form of
Ref.~\cite[Eq.~(31)]{KulLoeMerBaeKet2014}.

\subsubsection{Algorithmic Overview}

(Ai) We determine the parameters $I_{r:s}$, $M_{r:s}$, $V_{r:s}$, $\phi_{0}$,
Eq.~\eqref{eq:HrsActAng-all} as described in Ref.~\cite{EltSch2005}.

(Aii) We determine $\mathcal{H}_{0}(I)$, Eq.~\eqref{eq:H0}, by fitting it to
$N_{\text{disp}}$ tuples of action and frequency $(\bar{J},\bar{\omega})$
describing the tori of the regular region in the co-rotating frame of the
resonance.

(Aiii) We determine the near-identity transformations of
Eq.~\eqref{eq:CanTransIter}.
Initially, this requires sampling of the regular region using $N_{\text{ang}}$
points along $N_{\text{tori}}$ tori.
The invertibility of the near-identity transformations in a certain
phase-space region is ensured by rescaling the coefficients $a_{m,n}\mapsto
\eta a_{m,n}$ in Eq.~\eqref{eq:CanTransFamily} using a damping factor $\eta$.
If $N_{q}$, $N_{p}$ in Eq.~\eqref{eq:CanTransFamily} are too large, the tori of
the integrable approximation form curls and tendrils in the chaotic region.
In that case the integrable approximation cannot predict decay rates.
We control this problem by choosing the largest possible parameters $N_{q}$,
$N_{p}$ for which the tori of the integrable approximation provide a smooth
extrapolation into the chaotic phase-space region.
After a finite amount of steps $N_{\CanTrans}$, the canonical transformations
do not improve the agreement between the regular region and the integrable
approximation.
At this point we terminate the algorithm.

\subsubsection{Computational Parameters}
    \label{App:IntegrableApproximationListOfParameters}

In the following we list the important parameters of the integrable
approximation.

For $\K=2.9$ we use $I_{r:s}=0.009223$, $M_{r:s}=0.06243$, $V_{r:s}=-1.655 \cdot
10^{-7}$, and $\phi_{0}=\pi$.
For $\mathcal{H}_{0}$ in Eq.~\eqref{eq:H0} we used $N_{\text{disp}}=4$ and
fit its derivative to $N_{\text{disp}}=120$ tori of noble frequency.
We use $N_{\mathcal{T}}=40$ near-identity transformations,
Eq.~\eqref{eq:CanTransIter}, generated from Eq.~\eqref{eq:CanTransFamily} with
$N_{q}=N_{p}=2$ and coefficients rescaled by $\eta=0.05$.
The regular region was sampled using $N_{\text{ang}}=200$ points along
$N_{\text{tori}}=120$ tori, equidistantly distributed in action.

For $\K=3.4$ we use $I_{r:s}=0.01026$, $M_{r:s}=-0.047$, $V_{r:s}=-1.612 \cdot
10^{-5}$, and $\phi_{0}=0$.
For $\mathcal{H}_{0}$ in Eq.~\eqref{eq:H0} we use $N_{\text{disp}}=6$ and fit
its derivative to $N_{\text{disp}}=120$ tori, equidistantly distributed in
action.
We use $N_{\mathcal{T}}=15$ near-identity transformations,
Eq.~\eqref{eq:CanTransIter}, generated from Eq.~\eqref{eq:CanTransFamily} with
$N_{q}=N_{p}=2$ and coefficients rescaled by $\eta=0.25$.
The regular region is sampled using $N_{\text{ang}}=300$ points along
$N_{\text{tori}}=120$ tori, equidistantly distributed in action.

For $\K=3.5$ we use $I_{r:s}=0.01244$, $M_{r:s}=-0.048$, $V_{r:s}=-2.98 \cdot
10^{-5}$, and $\phi_{0}=0$.
For $\mathcal{H}_{0}$ in Eq.~\eqref{eq:H0} we use $N_{\text{disp}}=4$ and fit
its derivate to $N_{\text{disp}}=120$ tori, equidistantly distributed in
action.
We use $N_{\mathcal{T}}=15$ near-identity transformations,
Eq.~\eqref{eq:CanTransIter}, generated from Eq.~\eqref{eq:CanTransFamily} with
$N_{q}=N_{p}=2$ and coefficients rescaled by $\eta=0.25$.
The regular region is sampled using $N_{\text{ang}}=300$ points along
$N_{\text{tori}}=120$ tori, equidistantly distributed in action.

\subsubsection{Robustness}

\begin{figure}[tb!]
\begin{center}
\includegraphics[]{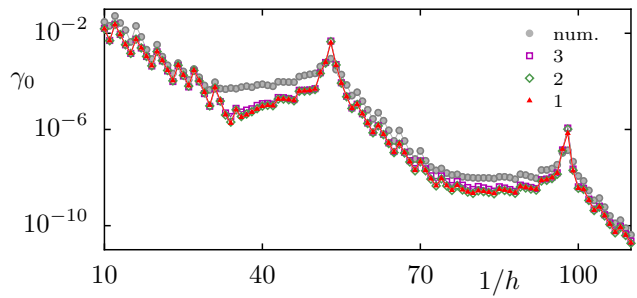}
  \caption{(color online) Decay rates $\gamma_{0}$ for the standard map at
$\K=3.4$ versus the inverse effective Planck constant.
Numerically determined rates ([gray] circles) are compared to predicted rates
according to Eq.~\eqref{eq:GammaPrediction} ([colored] symbols) based on three
slightly different integrable approximations.
}
     \label{fig:Robustness}
  \end{center}
\end{figure}

After fixing all parameters as described above the final integrable
approximation might differ, depending on the sampling of the regular region.
In order to show that this does not affect the final prediction, we evaluate
Eq.~\eqref{eq:GammaPrediction}, for three integrable approximations which are
based on slightly different sets of sample points.
The result is illustrated in Fig.~\ref{fig:Robustness}.
It shows that the prediction is clearly robust.

\subsection{Derivation of Quantization}
  \label{App:IntegrableApproximationQuantum}

In the following we sketch the basic ideas leading to the quantization
procedure presented in Sec.~\ref{Sec:Quantization}.
To this end we first present the quantization of the Hamilton-function obtained
after the transformation $\CanTrans^{0}$, Eqs.~\eqref{eq:CanTransInit} and
\eqref{eq:CanTransInitHO}, in Sec.~\ref{App:IntegrableApproximationQuantumT0}.
In Sec.~\ref{App:IntegrableApproximationQuantumT} we present how we extend
these results to the full transformation $\CanTrans$.

\subsubsection{Quantization after $\CanTrans^{0}$}
    \label{App:IntegrableApproximationQuantumT0}

To quantize the Hamilton-function $H_{r:s}^{(0)}(Q,P)$
obtained after the canonical transformation $\CanTrans^{0}$,
Eqs.~\eqref{eq:CanTransInit} and \eqref{eq:CanTransInitHO},
we follow Ref.~\cite{SchMouUll2011} by starting with the
transformed Hamilton-function
\begin{align}
 H_{r:s}^{(0)}(Q,P) &=   \mathcal{H}_{0}\left(\frac{Q^2 + P^2}{2}\right) +
\frac{\Vrs}{\left(2I_{r:s}\right)^{r/2}} \times \\ \nonumber
 & \left[\exp(\myi\phi_0) \left(\sigma^{1/2}[Q-q^{\star}] -
\myi\frac{P}{\sigma^{1/2}}\right)^r\right. \\ \nonumber
&\left.+ \exp(-\myi\phi_0) \left(\sigma^{1/2}[Q-q^{\star}]
+ \myi\frac{P}{\sigma^{1/2}}\right)^r \right].
\end{align}
In order to quantize this function we replace the coordinates $(Q,P)$ by
operators
\begin{subequations}
\begin{align}
 Q \mapsto \widehat{Q}\\
 P \mapsto \widehat{P}
\end{align}
\end{subequations}
and demand the usual commutation relation
\begin{align}
    \label{eq:CommutatrQP}
 [\widehat{Q}, \widehat{P}] = \myi\hbar.
\end{align}
This allows for introducing the corresponding ladder operators as
\begin{subequations}
  \label{eq:LadderoperatorQP}
  \begin{align}
\widehat{a} &:=
\frac{1}{(2\hbar)^{1/2}}\left(\sigma^{1/2}[\widehat{Q}-q^{\star}] +
\myi\frac{\widehat{P}}{\sigma^{1/2}}\right)\\
 \widehat{a}^{\dagger} &:=
\frac{1}{(2\hbar)^{1/2}}\left(\sigma^{1/2}[\widehat{Q}-q^{\star}] -
\myi\frac{\widehat{P}}{\sigma^{1/2}}\right)
  \end{align}
\end{subequations}
which admit the commutator
\begin{align}
 \label{eq:CommutatorLadderQP}
 [\widehat{a}, \widehat{a}^{\dagger}] = 1,
\end{align}
such that we get the number operator
\begin{align}
    \label{eq:numberOpQP}
 \widehat{n} := \widehat{a}^{\dagger}\widehat{a}.
\end{align}
Based on these operators the quantization of $H_{r:s}^{(0)}$ takes the form
\cite{SchMouUll2011}
\begin{align}
    \label{eq:HrsQP}
  \widehat{H}_{r:s}^{(0)} &=
\mathcal{H}_{0}(\widehat{I}) \\ \nonumber
&+ \Vrs\left(\frac{\hbar}{I_{r:s}}\right)^{\frac{r}{2}}
\left[\widehat{a}^{\dagger^{r}}\exp(\myi\phi_0) +
\widehat{a}^{r}\exp(-\myi\phi_0)\right],
\end{align}
where
\begin{align}
    \widehat{I} := \hbar(\widehat{n} + 1/2)
\end{align}
is the operator replacing the unperturbed action $I$.
Finally, in order to define the basis states, we identify them with the
eigenstates of the number operator leading to
\begin{align}
 \label{eq:eigenvalueNumberOpQP}
 \widehat{I}\Ket{I_n^{(0)}} &= I_n \Ket{I_n^{(0)}},
\end{align}
where the eigenvalues become quantizing actions $I_n=\hbar(n+1/2)$ and the basis
states $\Ket{I_n^{(0)}}$ fulfill
\begin{subequations}
\label{eq:basisstatesQP}
\begin{align}
 \widehat{a}\Ket{I_0^{(0)}} &= 0\\
 \Ket{I_n^{(0)}} &= \frac{1}{\sqrt{n!}}\widehat{a}^{\dagger^{n}}\Ket{I_0^{(0)}}.
\end{align}
\end{subequations}
With respect to this position basis $\Ket{I_n^{(0)}}$ become the eigenstates of
the harmonic oscillator
\begin{align}
 \label{eq:basisstatesQP_Qrepr}
 \!\!\!\BraKet{Q}{I_n^{(0)}} =
\left(\frac{\sigma}{\pi\hbar}\right)^{\frac{1}{4}} \!\!\!\!\frac{1}{\sqrt{2^n
n!}} \;\text{H}_{n}\!\left(\!\!\frac{Q}{\sqrt{\hbar/\sigma}}\!\!\right)
\exp{\!\left(\!\!-\frac{\sigma Q^2}{2\hbar}\right)},
\end{align}
where $\text{H}_{n}(\cdot)$ are the Hermite polynomials.

\subsubsection{Quantization after $\CanTrans$}
  \label{App:IntegrableApproximationQuantumT}

Our final goal is of course to obtain the quantization of the Hamilton-function
$H_{r:s}(q,p)$ which is related to $H_{r:s}^{(0)}(Q,P)$ via the canonical
transformation $\CanTrans'$, Eq.~\eqref{eq:CanTransIter}.
In order to obtain its quantization we assume that $\CanTrans'$
quantum-mechanically corresponds to a unitary operator
$\widehat{U}_{\CanTrans'}$ which has the following properties:
\begin{subequations}
\label{eq:UT}
\begin{align}
 \widehat{U}_{\CanTrans'}^{-1} &= \widehat{U}_{\CanTrans'^{-1}}\\
 \widehat{Q}' &=
\widehat{U}_{\CanTrans'}\widehat{Q}\widehat{U}_{\CanTrans'}^{-1}\\
 \widehat{P}' &=
\widehat{U}_{\CanTrans'}\widehat{P}\widehat{U}_{\CanTrans'}^{-1}
\end{align}
\end{subequations}
Such an operator exists at least within a semiclassical approximation
\cite{Bog1992}.
Note that $\widehat{Q}', \widehat{P}'$ represent the operators $\widehat{Q},
\widehat{P}$ within the final coordinate frame $(q,p)$.
However, they must not be confused with the operators $\widehat{q}, \widehat{p}$
which give rise to the position and momentum basis in the final coordinate
frame $(q,p)$.
In particular, while $\widehat{q}\ket{q}=q\ket{q}$, $\widehat{Q}\ket{q}\neq
q\ket{q}$.

Under the above assumption the transformed operators preserve the commutation
relation
\begin{align}
    \label{eq:Commutatrqp}
 [\widehat{Q}', \widehat{P}'] = \myi\hbar.
\end{align}
Furthermore, we get the transformed ladder operators as
\begin{subequations}
  \label{eq:Ladderoperatorqp}
  \begin{align}
\widehat{a}' &:=
\widehat{U}_{\CanTrans'}\widehat{a}\widehat{U}_{\CanTrans'}^{-1}\\
\widehat{a}'^{\dagger} &:=
\widehat{U}_{\CanTrans'}\widehat{a}^{\dagger}\widehat{U}_{\CanTrans'}^{-1}
  \end{align}
\end{subequations}
which admit the same commutator
\begin{align}
 \label{eq:CommutatorLadderqp}
 [\widehat{a}', \widehat{a}'^{\dagger}] = 1,
\end{align}
such that we get the transformed number operator
\begin{align}
    \label{eq:numberOpqp}
 \widehat{n}' =
\widehat{U}_{\CanTrans'}\widehat{n}\widehat{U}_{\CanTrans'}^{-1},
\end{align}
and the transformed action operator
\begin{align}
    \widehat{I}' := \hbar(\widehat{n}' + 1/2).
\end{align}

Based on these operators we can define the transformation of the quantization of
$H_{r:s}^{(0)}(Q,P)$ which we identify with the quantization of
$H_{r:s}(q,p)$.
It takes the form \cite{SchMouUll2011}
\begin{align}
    \label{eq:Hrsqp}
  \widehat{H}_{r:s} &=
\mathcal{H}_{0}(\widehat{I}') \\ \nonumber
&+ \Vrs\left(\frac{\hbar}{I_{r:s}}\right)^{\frac{r}{2}}
\left[\widehat{a}'^{\dagger^{r}}\exp(\myi\phi_0) +
\widehat{a}'^{r}\exp(-\myi\phi_0)\right].
\end{align}
Finally, in order to define the basis states $\ket{I_n}$, we identify them
with the eigenstates of the number operator $\widehat{n}'$, such that
\begin{align}
 \label{eq:eigenvalueNumberOpqp}
 \widehat{I}'\ket{I_n} &= I_n \ket{I_n}
\end{align}
with the basis states $\ket{I_n}$ which admit the property
\begin{subequations}
\label{eq:basisstatesqp}
\begin{align}
 \widehat{a}'\ket{I_0} &= 0\\
 \ket{I_n} &= \frac{1}{\sqrt{n!}}\widehat{a}'^{\dagger^{n}}\ket{I_0}.
\end{align}
\end{subequations}
Evaluating $\widehat{H}_{r:s}$, Eq.~\eqref{eq:Hrsqp} in the basis of
$\ket{I_n}$, based on Eqs.~\eqref{eq:basisstatesqp} gives the matrix
representation of Eq.~\eqref{eq:HrsActAngIBasis}.

Finally, for connecting $\widehat{H}_{r:s}$ and $\qmap$ we require the basis
states with respect to the basis $\ket{q}$.
To this end, one can show from the above equations that
\begin{align}
 \label{eq:basisstates_transformed_qp}
 \Ket{I_n} &= \widehat{U}_{\CanTrans'}\Ket{I_n^{(0)}},
\end{align}
such that
\begin{align}
 \label{eq:QuantumCanonicalTrafoOfStates}
 \BraKet{q}{I_n} = \int \text{d}Q \BraOpKet{q}{\widehat{U}_{\CanTrans'}}{Q}
\BraKet{Q}{I_n^{(0)}}.
\end{align}
In principle, the operator $\BraOpKet{q}{\widehat{U}_{\CanTrans'}}{q'}$ can be
evaluated semiclassically, using the techniques described in
Ref.~\cite{Bog1992}.
However, this does not give an analytical closed form result and its evaluation
is numerically extremely tedious.
Furthermore, $\widehat{U}_{\CanTrans'}$ is usually so close to an identity
transformation such that a semiclassical evaluation of
$\BraOpKet{q}{\widehat{U}_{\CanTrans'}}{Q}$ contains too many turning points.

Hence, we take an alternative approach, which is numerically feasible:\
(i) We recognize that the states $\ket{I_n}$ are the eigenstates of the operator
$\widehat{I}$, originating from the phase-space coordinate $I$.
(ii) We define the function $I(q,p)$ which is obtained after the full canonical
transformation $\CanTrans$.
(iii) We define the Weyl-quantization of this function on a phase-space torus
giving the hermitian matrix of Eq.~\eqref{eq:WeylI}.
(iv) We diagonalize this matrix numerically, yielding the states
$\BraKet{\overline{q}_l}{I_n}$.

Finally, obtaining the modes $\BraKet{\overline{q}_l}{I_n}$ from an eigenvalue
equation comes at the cost that their relative phase (usually ensured via
Eq.~\eqref{eq:basisstatesqp} or alternatively via Eqs.~\eqref{eq:basisstatesQP}
and \eqref{eq:basisstates_transformed_qp}) is lost.
For the standard map we try to restore this phase by exploiting the symmetry of
Eq.~\eqref{eq:WeylI}, which for our system becomes a real symmetric matrix.
In that we can ensure that the coefficient vector $\BraKet{\overline{q}_l}{I_n}$
can be chosen real.
Finally, we fix the sign of this coefficient vector, by aligning
it with the mode $\BraKet{Q}{I_n^{(0)}}$ defined via
Eq.~\eqref{eq:basisstatesQP_Qrepr}.
This means, we choose the sign of the coefficient vector
$\BraKet{\overline{q}_l}{I_n}$ such that the following relation is
fulfilled
\begin{align}
 \label{eq:ModeAligning}
 \sum_{n} \BraKet{I_n}{\overline{q}_l}
\left.\left[\BraKet{Q}{I_n^{(0)}}\right]\right|_{Q=\overline{q}_l} > 0.
\end{align}
This assumes that the unitary operator representing the quantum
canonical transformation in Eq.~\eqref{eq:QuantumCanonicalTrafoOfStates} is
sufficiently close to an identity transformation
$\widehat{U}_{\CanTrans'}\approx1$.

\end{appendix}


\begin{thebibliography}{10}
\newcommand{\enquote}[1]{``#1''}
% \providecommand{\url}[1]{\texttt{#1}}
% \providecommand{\urlprefix}{URL }
% \providecommand{\eprint}[2][]{\url{#2}}

\bibitem{DavHel1981}
M.~J. Davis and E.~J. Heller, \emph{Quantum dynamical tunneling in bound
  states}, J.~Chem.~Phys. \textbf{75}, 246 (1981).

\bibitem{KesSch2011}
S.~Keshavamurthy and P.~Schlagheck, \emph{Dynamical Tunneling: Theory and
  Experiment}, Taylor \& Francis, Boca Raton (2011).

\bibitem{LinBal1990}
W.~A. Lin and L.~E. Ballentine, \emph{Quantum tunneling and chaos in a driven
  anharmonic oscillator}, Phys.~Rev.~Lett. \textbf{65}, 2927 (1990).

\bibitem{BohTomUll1993}
O.~Bohigas, S.~Tomsovic, and D.~Ullmo, \emph{Manifestations of classical phase
  space structures in quantum mechanics}, Phys.~Rep. \textbf{223}, 43 (1993).

\bibitem{ShiHarFukHenSasNar2010}
S.~Shinohara, T.~Harayama, T.~Fukushima, M.~Hentschel, T.~Sasaki, and E.~E.
  Narimanov, \emph{Chaos-assisted directional light emission from microcavity
  lasers}, Phys.~Rev.~Lett. \textbf{104}, 163902 (2010).

\bibitem{ShiHarFukHenSunNar2011}
S.~Shinohara, T.~Harayama, T.~Fukushima, M.~Hentschel, S.~Sunada, and E.~E.
  Narimanov, \emph{Chaos-assisted emission from asymmetric resonant cavity
  microlasers}, Phys.~Rev.~A \textbf{83}, 053837 (2011).

\bibitem{YanLeeMooLeeKimDaoLeeAn2010}
J.~Yang, S.-B. Lee, S.~Moon, S.-Y. Lee, S.~W. Kim, T.~T.~A. Dao, J.-H. Lee, and
  K.~An, \emph{Pump-induced dynamical tunneling in a deformed microcavity
  laser}, Phys.~Rev.~Lett. \textbf{104}, 243601 (2010).

\bibitem{KwaShiMooLeeYanAn2015}
H.~Kwak, Y.~Shin, S.~Moon, S.-B. Lee, J.~Yang, and K.~An, \emph{Nonlinear
  resonance-assisted tunneling induced by microcavity deformation}, Scientific
  Reports \textbf{5}, 9010 (2015).

\bibitem{YiYuLeeKim2015}
C.-H. Yi, H.-H. Yu, J.-W. Lee, and C.-M. Kim, \emph{Fermi resonance in optical
  microcavities}, Phys.~Rev.~E \textbf{91}, 042903 (2015).

\bibitem{YiYuKim2016}
C.-H. Yi, H.-H. Yu, and C.-M. Kim, \emph{Resonant torus-assisted tunneling},
  Phys.~Rev.~E \textbf{93}, 012201 (2016).

\bibitem{DemGraHeiHofRehRic2000}
C.~Dembowski, H.-D. Gr\"af, A.~Heine, R.~Hofferbert, H.~Rehfeld, and
  A.~Richter, \emph{First experimental evidence for chaos-assisted tunneling in
  a microwave annular billiard}, Phys.~Rev.~Lett. \textbf{84}, 867 (2000).

\bibitem{BaeKetLoeRobVidHoeKuhSto2008}
A.~B\"acker, R.~Ketzmerick, S.~L\"ock, M.~Robnik, G.~Vidmar, R.~H\"ohmann,
  U.~Kuhl, and H.-J. St\"ockmann, \emph{Dynamical tunneling in mushroom
  billiards}, Phys.~Rev.~Lett. \textbf{100}, 174103 (2008).

\bibitem{DieGuhGutMisRic2014}
B.~Dietz, T.~Guhr, B.~Gutkin, M.~Miski-Oglu, and A.~Richter, \emph{Spectral
  properties and dynamical tunneling in constant-width billiards}, Phys.~Rev.~E
  \textbf{90}, 022903 (2014).

\bibitem{GehLoeShiBaeKetKuhSto2015}
S.~Gehler, S.~L{\"o}ck, S.~Shinohara, A.~B{\"a}cker, R.~Ketzmerick, U.~Kuhl,
  and H.-J. St{\"o}ckmann, \emph{Experimental observation of resonance-assisted
  tunneling}, Phys.~Rev.~Lett. \textbf{115}, 104101 (2015).

\bibitem{Hen2001}
W.~K. {Hensinger et al.}, \emph{Dynamical tunnelling of ultracold atoms},
  Nature \textbf{412}, 52 (2001).

\bibitem{SteOskRai2001}
D.~A. Steck, W.~H. Oskay, and M.~G. Raizen, \emph{Observation of chaos-assisted
  tunneling between islands of stability}, Science \textbf{293}, 274 (2001).

\bibitem{UzeNoiMar1983}
T.~Uzer, D.~W. Noid, and R.~A. Marcus, \emph{Uniform semiclassical theory of
  avoided crossings}, J.~Chem.~Phys. \textbf{79}, 4412 (1983).

\bibitem{Ozo1984}
A.~M. Ozorio~de Almeida, \emph{Tunneling and the semiclassical spectrum for an
  isolated classical resonance}, J.~Phys.~Chem. \textbf{88}, 6139 (1984).

\bibitem{BroSchUll2001}
O.~Brodier, P.~Schlagheck, and D.~Ullmo, \emph{Resonance-assisted tunneling in
  near-integrable systems}, Phys.~Rev.~Lett. \textbf{87}, 064101 (2001).

\bibitem{BroSchUll2002}
O.~Brodier, P.~Schlagheck, and D.~Ullmo, \emph{Resonance-assisted tunneling},
  Ann.~Phys.~(N.Y.) \textbf{300}, 88 (2002).

\bibitem{ZakDelBuc1998}
J.~Zakrzewski, D.~Delande, and A.~Buchleitner, \emph{Ionization via chaos
  assisted tunneling}, Phys.~Rev.~E \textbf{57}, 1458 (1998).

\bibitem{BucDelZak2002}
A.~Buchleitner, D.~Delande, and J.~Zakrzewski, \emph{Non-dispersive wave
  packets in periodically driven quantum systems}, Phys.~Rep. \textbf{368}, 409
  (2002).

\bibitem{WimSchEltBuc2006}
S.~Wimberger, P.~Schlagheck, C.~Eltschka, and A.~Buchleitner,
  \emph{Resonance-assisted decay of nondispersive wave packets},
  Phys.~Rev.~Lett. \textbf{97}, 043001 (2006).

\bibitem{ShuIke1995}
A.~Shudo and K.~S. Ikeda, \emph{Complex classical trajectories and chaotic
  tunneling}, Phys.~Rev.~Lett. \textbf{74}, 682 (1995).

\bibitem{ShuIke1998}
A.~Shudo and K.~S. Ikeda, \emph{Chaotic tunneling: A remarkable manifestation
  of complex classical dynamics in non--integrable quantum phenomena},
  Physica~D \textbf{115}, 234 (1998).

\bibitem{PodNar2003}
V.~A. Podolskiy and E.~E. Narimanov, \emph{Semiclassical description of
  chaos-assisted tunneling}, Phys.~Rev.~Lett. \textbf{91}, 263601 (2003).

\bibitem{PodNar2005}
V.~A. Podolskiy and E.~E. Narimanov, \emph{Chaos-assisted tunneling in
  dielectric microcavities}, Opt.~Lett. \textbf{30}, 474 (2005).

\bibitem{EltSch2005}
C.~Eltschka and P.~Schlagheck, \emph{Resonance- and chaos-assisted tunneling in
  mixed regular-chaotic systems}, Phys.~Rev.~Lett. \textbf{94}, 014101 (2005).

\bibitem{SheFisGuaReb2006}
M.~Sheinman, S.~Fishman, I.~Guarneri, and L.~Rebuzzini, \emph{Decay of quantum
  accelerator modes}, Phys.~Rev.~A \textbf{73}, 052110 (2006).

\bibitem{BaeKetLoeSch2008}
A.~B\"acker, R.~Ketzmerick, S.~L\"ock, and L.~Schilling,
  \emph{Regular-to-chaotic tunneling rates using a fictitious integrable
  system}, Phys.~Rev.~Lett. \textbf{100}, 104101 (2008).

\bibitem{ShuIshIke2008}
A.~Shudo, Y.~Ishii, and K.~S. Ikeda, \emph{Chaos attracts tunneling
  trajectories: A universal mechanism of chaotic tunneling}, Europhys.~Lett.
  \textbf{81}, 50003 (2008), 00013.

\bibitem{ShuIshIke2009a}
A.~Shudo, Y.~Ishii, and K.~S. Ikeda, \emph{Julia sets and chaotic tunneling:
  I}, J.~Phys.~A \textbf{42}, 265101 (2009).

\bibitem{ShuIshIke2009b}
A.~Shudo, Y.~Ishii, and K.~S. Ikeda, \emph{Julia sets and chaotic tunneling:
  {II}}, J.~Phys.~A \textbf{42}, 265102 (2009).

\bibitem{BaeKetLoeWieHen2009}
A.~B\"acker, R.~Ketzmerick, S.~L\"ock, J.~Wiersig, and M.~Hentschel,
  \emph{Quality factors and dynamical tunneling in annular microcavities},
  Phys.~Rev.~A \textbf{79}, 063804 (2009).

\bibitem{BaeKetLoe2010}
A.~B\"acker, R.~Ketzmerick, and S.~L\"ock, \emph{Direct regular-to-chaotic
  tunneling rates using the fictitious-integrable-system approach},
  Phys.~Rev.~E \textbf{82}, 056208 (2010).

\bibitem{LoeBaeKetSch2010}
S.~L\"ock, A.~B\"acker, R.~Ketzmerick, and P.~Schlagheck,
  \emph{Regular-to-chaotic tunneling rates: From the quantum to the
  semiclassical regime}, Phys.~Rev.~Lett. \textbf{104}, 114101 (2010).

\bibitem{MerLoeBaeKetShu2013}
N.~Mertig, S.~L\"ock, A.~B\"acker, R.~Ketzmerick, and A.~Shudo, \emph{Complex
  paths for regular-to-chaotic tunnelling rates}, Europhys.~Lett. \textbf{102},
  10005 (2013).

\bibitem{HanShuIke2015}
Y.~Hanada, A.~Shudo, and K.~S. Ikeda, \emph{Origin of the enhancement of
  tunneling probability in the nearly integrable system}, Phys.~Rev.~E
  \textbf{91}, 042913 (2015).

\bibitem{KulWie2016}
J.~Kullig and J.~Wiersig, \emph{Q spoiling in deformed optical microdisks due
  to resonance-assisted tunneling}, \emph{accepted for publication in} Phys. 
  Rev. E.
  
\bibitem{SchMouUll2011}
P.~Schlagheck, A.~Mouchet, and D.~Ullmo, \emph{Resonance-assisted tunneling in
  mixed regular-chaotic systems}, in \enquote{Dynamical Tunneling: Theory and
  Experiment},  \cite{KesSch2011}, chapter~8, 177.

\bibitem{DeuMou2010}
J.~Le~Deunff and A.~Mouchet, \emph{Instantons re-examined: Dynamical tunneling
  and resonant tunneling}, Phys.~Rev.~E \textbf{81}, 046205 (2010).

\bibitem{DeuMouSch2013}
J.~Le~Deunff, A.~Mouchet, and P.~Schlagheck, \emph{Semiclassical description of
  resonance-assisted tunneling in one-dimensional integrable models},
  Phys.~Rev.~E \textbf{88}, 042927 (2013).

\bibitem{KulLoeMerBaeKet2014}
J.~Kullig, C.~L\"{o}bner, N.~Mertig, A.~B\"{a}cker, and R.~Ketzmerick,
  \emph{Integrable approximation of regular regions with a nonlinear resonance
  chain}, Phys.~Rev.~E \textbf{90}, 052906 (2014).

\bibitem{FriMerLoeBaeKet}
F.~Fritzsch, N.~Mertig, C.~L\"{o}bner, A.~B\"{a}cker, and R.~Ketzmerick,
  \emph{in preparation}.

\bibitem{Chi1979}
B.~V. {Chirikov}, \emph{{A universal instability of many-dimensional oscillator
  systems}}, Phys.~Rep. \textbf{52}, 263 (1979).

\bibitem{Kol1954}
A.~N. Kolmogorov, \emph{Preservation of conditionally periodic movements with
  small change in the {H}amilton function \rm (in {Russian})},
  Dokl.~Akad.~Nauk.~SSSR \textbf{98}, 527 (1954), english translation in
  \cite{CasFor1979}, 51--56.

\bibitem{Arn1963}
V.~I. Arnold, \emph{Small denominators and problems of stability of motion in
  classical and celestial mechanics}, Russ.~Math.~Surv. \textbf{18}, 85 (1963).

\bibitem{Arn1963b}
V.~I. Arnold, \emph{Proof of a theorem of {A}.~{N}.~{K}olmogorov on the
  invariance of quasi-periodic motions under small perturbations of the
  {H}amiltonian}, Russ.~Math.~Surv. \textbf{18}, 9 (1963).

\bibitem{Mos1962}
J.~Moser, \emph{On invariant curves of area-preserving mappings of an annulus},
  Nachr. Akad. Wiss. G\"ottingen \textbf{1}, 1 (1962).

\bibitem{Bir1913}
G.~D. Birkhoff, \emph{Proof of {P}oincar\'e's geometric theorem},
  Trans.~Amer.~Math.~Soc. \textbf{14}, 14 (1913).

\bibitem{LicLie1983a}
A.~J. Lichtenberg and M.~A. Lieberman, \emph{Regular and Chaotic Motion},
  volume~38 of \emph{Applied Mathematical Sciences}, {S}pringer-{V}erlag, {N}ew
  {Y}ork, 2 edition (1992).

\bibitem{BerBalTabVor1979}
M.~V. Berry, N.~L. Balazs, M.~Tabor, and A.~Voros, \emph{Quantum maps},
  Ann.~Phys.~(N.Y.) \textbf{122}, 26 (1979).

\bibitem{HanBer1980}
J.~H. Hannay and M.~V. Berry, \emph{Quantization of linear maps on a torus ---
  {F}resnel diffraction by a periodic grating}, Physica~D \textbf{1}, 267
  (1980).

\bibitem{ChaShi1986}
S.-J. Chang and K.-J. Shi, \emph{Evolution and exact eigenstates of a resonant
  quantum system}, Phys.~Rev.~A \textbf{34}, 7 (1986).

\bibitem{KeaMezRob1999}
J.~P. Keating, F.~Mezzadri, and J.~M. Robbins, \emph{Quantum boundary
  conditions for torus maps}, Nonlinearity \textbf{12}, 579 (1999).

\bibitem{DegGra2003b}
M.~Degli~Esposti and S.~Graffi, \emph{Mathematical aspects of quantum maps}, in
  \enquote{The Mathematical Aspects of Quantum Maps},  \cite{DegGra2003}, 49.

\bibitem{Bae2003}
A.~B\"acker, \emph{Numerical aspects of eigenvalues and eigenfunctions of
  chaotic quantum systems}, in Degli~Esposti and Graffi  \cite{DegGra2003}, 91.

\bibitem{Per1973}
I.~C. Percival, \emph{Regular and irregular spectra}, J.~Phys.~B \textbf{6},
  L229 (1973).

\bibitem{Ber1977b}
M.~V. Berry, \emph{Regular and irregular semiclassical wavefunctions},
  J.~Phys.~A \textbf{10}, 2083 (1977).

\bibitem{Vor1979}
A.~Voros, \emph{Semi-classical ergodicity of quantum eigenstates in the
  {W}igner representation}, in Casati and Ford  \cite{CasFor1979}, 326.

\bibitem{Boh1913}
N.~Bohr, \emph{On the constitution of atoms and molecules}, Philosophical
  Magazine Series 6 \textbf{26}, 1 (1913).

\bibitem{Boh1913b}
N.~Bohr, \emph{On the constitution of atoms and molecules. {P}art {II}. --
  {S}ystems containing only a single nucleus}, Philosophical Magazine Series 6
  \textbf{26}, 476 (1913).

\bibitem{Som1916}
A.~Sommerfeld, \emph{Zur {Q}uantentheorie der {S}pektrallinien}, Ann.~Phys.
  \textbf{356}, 1 (1916).

\bibitem{WisSarArrBenBor2011}
D.~A. Wisniacki, M.~Saraceno, F.~J. Arranz, R.~M. Benito, and F.~Borondo,
  \emph{Poincar\'{e}-{B}irkhoff theorem in quantum mechanics}, Phys.~Rev.~E
  \textbf{84}, 026206 (2011).

\bibitem{Wis2014}
D.~A. Wisniacki, \emph{Universal wave functions structure in mixed systems},
  Europhys.~Lett. \textbf{106}, 60006 (2014).

\bibitem{WisSch2015}
D.~A. Wisniacki and P.~Schlagheck, \emph{Quantum manifestations of classical
  nonlinear resonances}, Phys.~Rev.~E \textbf{92}, 062923 (2015).

\bibitem{HanOttAnt1984}
J.~D. Hanson, E.~Ott, and T.~M. Antonsen, \emph{Influence of finite wavelength
  on the quantum kicked rotator in the semiclassical regime}, Phys.~Rev.~A
  \textbf{29}, 819 (1984).

\bibitem{LoeLoeBaeKet2013}
C.~L\"obner, S.~L\"ock, A.~B\"acker, and R.~Ketzmerick, \emph{Integrable
  approximation of regular islands: The iterative canonical transformation
  method}, Phys.~Rev.~E \textbf{88}, 062901 (2013).

\bibitem{KayMeiPer1984a}
R.~S. MacKay, J.~D. Meiss, and I.~C. Percival, \emph{Stochasticity and
  transport in {Hamiltonian} systems}, Phys.~Rev.~Lett. \textbf{52}, 697
  (1984).

\bibitem{KayMeiPer1984b}
R.~S. MacKay, J.~D. Meiss, and I.~C. Percival, \emph{{Transport in
  {H}amiltonian systems}}, Physica~D \textbf{13}, 55 (1984).

\bibitem{Bog1992}
E.~B. Bogomolny, \emph{{Semiclassical quantization of multidimensional
  systems}}, Nonlinearity \textbf{5}, 805 (1992).

\bibitem{CasFor1979}
G.~Casati and J.~Ford (editors) \emph{Stochastic Behavior in Classical and
  Quantum {Hamiltonian} Systems}, volume~93 of \emph{Lect.~Notes Phys.},
  Springer-Verlag, Berlin (1979).

\bibitem{DegGra2003}
M.~Degli~Esposti and S.~Graffi (editors) \emph{The Mathematical Aspects of
  Quantum Maps}, volume 618 of \emph{Lect.~Notes Phys.}, Springer-Verlag,
  Berlin (2003).

\end{thebibliography}
\end{document}